\documentclass[journal,comsoc]{IEEEtran}
\usepackage{graphicx}
\usepackage{array}
\usepackage{color}
\usepackage{epsf}
\usepackage{times}
\usepackage{epsfig}
\usepackage{graphicx}
\usepackage{epstopdf}
\usepackage{hyperref}
\usepackage{cite}
\usepackage{amsmath}
\usepackage{amssymb}
\usepackage{amsxtra}
\usepackage{amsthm}
\usepackage{bbm}
\usepackage{algorithmic}
\usepackage{algorithm}
\usepackage{subfigure}
\usepackage{authblk}
\usepackage{url}
\usepackage{comment}
\usepackage{siunitx}
\usepackage{lipsum}
\usepackage{kantlipsum}
\usepackage{dblfloatfix}
\usepackage{adjustbox}
\usepackage{tabularx}
\usepackage{multirow}
\usepackage{float}
\usepackage{framed}
\usepackage{caption}

\graphicspath{ {Figures/} }

\begin{document}

\title{On the Vulnerability of Underwater Magnetic Induction Communication 
}

\author{Muhammad Muzzammil, Waqas Aman, Irfan Ullah, Shang Zhigang, Saif Al-Kuwari,~\IEEEmembership{Senior Member,~IEEE}, Zhou Tian, Marwa Qaraqe,~\IEEEmembership{Senior Member,~IEEE} 

\thanks{ This work is partially funded by the G5828 ``SeaSec: DroNets for Maritime Border and Port Security" project under the NATO's Science for Peace Programme.  }
\thanks{Muhammad Muzzammil, Irfan Ullah, Shang Zhigang, and Zhou Tian are with the Acoustic Science and Technology Laboratory, Harbin Engineering University, Harbin 150001, China, the Key Laboratory of Marine Information Acquisition and Security, Harbin Engineering University, Ministry of Industry and Information Technology, Harbin 150001, China, and the College of Underwater Acoustic Engineering, Harbin Engineering University, China, Email: \{muzzammilm, irfankhan, shangzhigang, zhoutian\}@hrbeu.edu.cn.\newline
W. Aman, S. Al-Kuwari and M. Qaraqe are with the Division of Information and Computing Technology, College of Science and Engineering, Hamad Bin Khalifa University, Doha, Qatar, Emails: \{waman, smalkuwari, mqaraqe\}@hbku.edu.qa.\\
\textbf{Corresponding Author:} Muhammad Muzzammil, Email: muzzammilm@hrbeu.edu.cn
}

}

\maketitle

\begin{abstract} 
Typical magnetic induction (MI) communication is commonly considered a secure underwater wireless communication (UWC) technology due to its non-audible and non-visible nature compared to acoustic and optical UWC technologies. However, vulnerabilities in communication systems inevitably exist and may lead to different types of attacks. In this paper, we investigate the eavesdropping attack in underwater MI communication to quantitatively measure the system's vulnerability under this attack. We consider different potential eavesdropping configuration setups based on the positions and orientations of the eavesdropper node to investigate how they impact the received voltage and secrecy at the legitimate receiver node. To this end, we develop finite-element-method-based simulation models for each configuration in an underwater environment and evaluate the received voltage and the secrecy capacity against different system parameters such as magnetic flux, magnetic flux density, distance, and orientation sensitivity. Furthermore, we construct an experimental setup within a laboratory environment to replicate the simulation experiments. Both simulation and lab experimental confirm the susceptibility of underwater MI communication to eavesdropping attacks. However, this vulnerability is highly dependent on the position and orientation of the coil between the eavesdropper and the legitimate transmitter. On the positive side, we also observe a unique behavior in the received coil reception that might be used to detect malicious node activities in the vicinity, which might lead to a potential security mechanism against eavesdropping attacks. 

\end{abstract}

\begin{keywords}
Magnetic induction, eavesdropping, vulnerability,  underwater, distance, orientation sensitivity, secrecy capacity, received power, and finite element method.
\end{keywords}

\section{Introduction}
\label{sec:intro}

Underwater wireless communication (UWC) technologies such as acoustic, optics, electromagnetic, and magnetic induction (MI) facilitate the exploration of vast oceanic systems, which are rich in plentiful natural resources such as oil and gas, and play a significant role in essential applications, such as combating climate change and underwater monitoring~\cite{li2019survey, muzzammil2020fundamentals}. In addition, UWC technologies are also widely used in military and surveillance applications \cite{ali2020recent,muzzammil2022towards}. The choice of UWC technologies should be application-dependent, as the harsh nature of the underwater environment greatly affects communication channels, which have unique characteristics according to the underlying technology \cite{gussen2016survey,aman2023security}.

Acoustic communication is widely used in underwater environments due to the long propagation nature of acoustic waves, making it the most feasible option for long-distance communication. However, it faces significant challenges such as low data rate, high link delay, large multipath, and the Doppler effect \cite{muzzammil2020fundamentals, wan2020fine}. On the other hand, underwater optical wireless communication can be ideal for high data rates, low link delay, and medium-range communication, but it experiences high attenuation, scattering, and line of sight requirements \cite{zeng2017survey}. Magnetic induction (MI) communication has recently emerged as a promising alternative for medium-range underwater applications, offering stable channel response, absence of multipath and Doppler effects, and smooth cross-boundary communication due to the same magnetic permeability in air and water \cite{Akyildiz2015Realizing, muzzammil2020fundamentals}. However, MI communication faces challenges related to the orientation of the coil and the conductive nature of water \cite{muzzammil2019directivity}.

UWC technologies suffer from intrinsic vulnerabilities due to their broadcast nature. This makes UWC (like other wireless technologies) susceptible to conventional wireless communication attacks, including eavesdropping \cite{aman2020maximizing} and jamming \cite{aman2023security}. In the literature, numerous studies have investigated the vulnerabilities of underwater acoustic communication~\cite{wang2016modeling, jiang2018securing, ahmad2021analysis, aman2023security}. Similarly, vulnerabilities in underwater optical communication have also been reported~\cite{kong2017security, zhang2023security, boluda2023impact}. However, studies on the vulnerabilities of underwater magnetic induction communication are quite rare and are mainly focused on short-range near-field communication (NFC)~\cite{nelson2013security, rahman2017classification, sun2017magnetic, al2017towards, singh2018near, alrobaish2022common}. 

In \cite{nelson2013security}, the authors summarize various security attacks on NFC communication. A similar study is reported in \cite{rahman2017classification}, which provides a classification of near-field communication attacks and discusses security concerns in NFC by targeting mainly five short-range applications such as healthcare and e-payment.  \cite{singh2018near} presented the vulnerabilities in NFC and focused on the denial of service attack and data corruption attack. In \cite{alrobaish2022common}, the authors studied broken access control attacks, especially in e-payment services that use NFC technology. 
Security in wireless power transfer applications based on magnetic coupling is studied in \cite{sun2017magnetic}, and countermeasure techniques for eavesdropping in NFC are studied in \cite{al2017towards}. In \cite{han2018secure}, the authors studied the secrecy capacity of a simultaneous wireless information and power transfer system based on magnetic inductive coupling. Since these studies are based on NFC with a range of up to $\approx 10cm$, their target applications are mainly healthcare, e-payment, and tracking. 
However, 
in this paper, we quantitatively study the vulnerabilities of MI-based underwater wireless communication against eavesdropping attacks and show that MI is indeed susceptible to such attacks. The main contributions made in this paper are summarized in following subsection.

\subsection{Contribution }
In this paper, we investigate the eavesdropping attack on underwater MI communication. In a typical eavesdropping scenario, the eavesdropper node can receive a magnetic field and, when approaching the vicinity of legitimate ongoing MI communication, can listen to the legitimate communication (see Section \ref{sec:eaves}). Hence, we consider an underwater environment with three nodes: one legitimate transmitter (Tx) node, one legitimate receiver (Rx) node, and one eavesdropper node. The contributions of this paper can be summarized as follows: 

\begin{itemize}
    \item We analyze a system model for an eavesdropping attack on legitimate underwater MI communication by considering two main challenges: the placement and orientation sensitivity of the eavesdropper node with respect to the legitimate nodes. We attempt to address two main questions: \textit{First}, whether and how much information an eavesdropper node can receive from an ongoing legitimate MI communication. \textit{Second}, whether legitimate nodes can detect malicious activity in the environment.
    \item We develop various configurations to assess the impact of the position and orientation of the eavesdropper node on legitimate nodes. In the position case, we use two different configurations based on the far and near position of an eavesdropper node with respect to legitimate nodes. In the orientation case, we use three different configurations; changing the orientation of an eavesdropper node in a rotational fashion with respect to legitimate Tx, Rx, and around its own origin. 
    We develop a finite element method (FEM) simulation for underwater MI communication models (based on various configurations mentioned earlier) to quantitatively measure the vulnerability (Section \ref{sec:sim-results}).
    \item To verify the FEM simulation results, we further conduct lab experiments (Section \ref{sec:exp}). 
\end{itemize}


\subsection{Organization}
The rest of this paper is organized as follows: Section \ref{sec:preliminaries} provides a brief background on MI communication and eavesdropping attack. The evaluation of our system model using FEM simulation is presented in Section \ref{sec:sim-results} followed by experimental validation in Section \ref{sec:exp}. Finally, the paper concludes in Section \ref{sec:conclusion} with a few concluding remarks and future directions.



\section{Preliminaries}
\label{sec:preliminaries}
\subsection{MI Communication}
\label{sec:backgroundMI}

 MI communication works as follows: a pair of low-cost wounded coils are used, where one coil (Tx) is excited with a time-varying signal to generate a time-varying magnetic field $B$, given by \cite{ravindran2018characterization, hott2020underwater}

\begin{equation}
\label{eq:mfield}
    B = \frac{\mu_0 I_{Tx} N r^2 cos(\theta)}{2(r^2+d^2)^{3/2}},
\end{equation}
where $\mu_0=4\pi \times 10^{-7}$ is the permeability constant, $N$ is the number of turns, $I_{Tx}$ is the time-varying current supplied to the Tx coil, $r$ is the radius of the coil, $d$ is the distance from the Tx coil, and $\theta$ is the angle between the Tx and Rx coils. Eq. \eqref{eq:mfield} can be used in air medium that is a non-conducting medium, however, the magnetic field signals can be further attenuated due to eddy current in a conducting medium such as seawater. The magnetic field therefore becomes \cite{hott2020underwater}

\begin{equation} 
B = \frac {\mu _{0} N_{\mathrm w} I_{Tx} r^{2} \cos {(\theta)}}{2(r^{2} + d^{2})^{3/2}} \, e^{-\frac {\sqrt {d^{2}+r^{2}}}{\delta }},
\end{equation}
where $\delta=\frac {1}{\sqrt {\pi f \mu \sigma }}$ represents the skin depth (in the case of good conductors) and $\sigma$ is the conductivity of the medium. The receiver coil couples with the Tx coil when it comes closer to it and voltage is induced ($V_{ind}$) to the Rx coil, which can be expressed as \cite{ravindran2018characterization, ahmed2018design}

\begin{equation}
\label{eq:IV}
    V_{ind} = 2 \pi N A B cos (\theta),
\end{equation}
where $N$ and $A$ denote the number of turns and area of the coil, $B$ is the magnetic field strength, and $\theta$ is the angle of arrival. Therefore, the two coils coupled with each other, and this coupling is equal to $k=M/\sqrt{L_{Tx}L_{Rx}}$ \cite{agbinya2012power}, where $L_{Tx}$ and $L_{Rx}$ denote the inductance values of Tx and Rx coils, while $M$ is the mutual inductance between the two coils and can be expressed as \cite{muzzammil2020fundamentals, domingo2012magnetic}

\begin{equation}
        M={\mu_r \mu_0 \pi  N_{Tx} r_{Tx}^{2} N_{Rx} r_{Rx}^{2}\over 2\sqrt{\left(r_{Tx}^{2}+d^{2}\right)^{3}}},  \label{eq:mutual_inductance}
\end{equation}
where $\mu_o=4\pi \times 10^7 H/m$ is the magnetic permeability constant and $\mu_r=1$ is the relative permeability of water; $N_{Tx}$ and $ N_{Rx}$ are the number of turns of the Tx and Rx coil, $r_{Tx}$ and $ r_{Rx}$ is the radius of the Tx and Rx coil and $d$ is the distance between Tx and Rx coil.

The coupling strength of the Tx-Rx coils can be described in many terms, such as the value of the coupling coefficient (range between 0 and 1) or the received voltage at the receiver. A coupling coefficient with a value 1 means that Tx and Rx are strongly coupled, while a value 0 means that they are no longer coupled. Similarly, the maximum voltage received at the Rx coil means strong coupling and vice versa. Furthermore, strong coupling depends on many factors, such as how close the coils are to each other and the magnetic moment, which is $m=NIAcos(\theta)$, where $N$, $A$, and $I$ denote the number of turns, the area of the coil, and the current supplied to the Tx coil. $\theta$ represents the angle between the Tx and Rx coil. 

The communication range through MI directly depends on the magnetic moment; increasing magnetic leads to higher received magnetic field strength, which facilitates longer communication distance and vice versa. Since the attenuation rate of the magnetic field strength decays very rapidly ($1/d^3$, where $d$ is the communication distance), one of the major challenges in MI communication is the limited range.

The angle term in the magnetic moment introduces an important challenge in MI communication; that is, when $\theta=0^\circ$, Tx and Rx coils are perfectly aligned, meaning that Rx receives the maximum magnetic field strength. On the other hand, if $\theta \neq 0^\circ$, Tx and Rx coils are not aligned and Rx consequently receives a weak magnetic field. Therefore, the coil's orientation between any two MI communicating nodes plays a significant role in MI communication. In the following sections, we focus on these main challenges of MI communication in our study of eavesdropping attacks.



\subsection{Eavesdropping}
\label{sec:eaves}
An eavesdropping attack, also known as passive wiretapping or snooping, is a form of unauthorized interception of communications between two parties. This attack involves an adversary secretly monitoring and listening to conversations, data transmissions, or any form of communication between intended participants without their knowledge or consent. An eavesdropping attack is therefore considered a breach of privacy and security and can have serious consequences, such as compromise of critical infrastructure, identity, and data theft. 

\begin{figure}[t]
    \centering
    \includegraphics[width=0.9\linewidth]{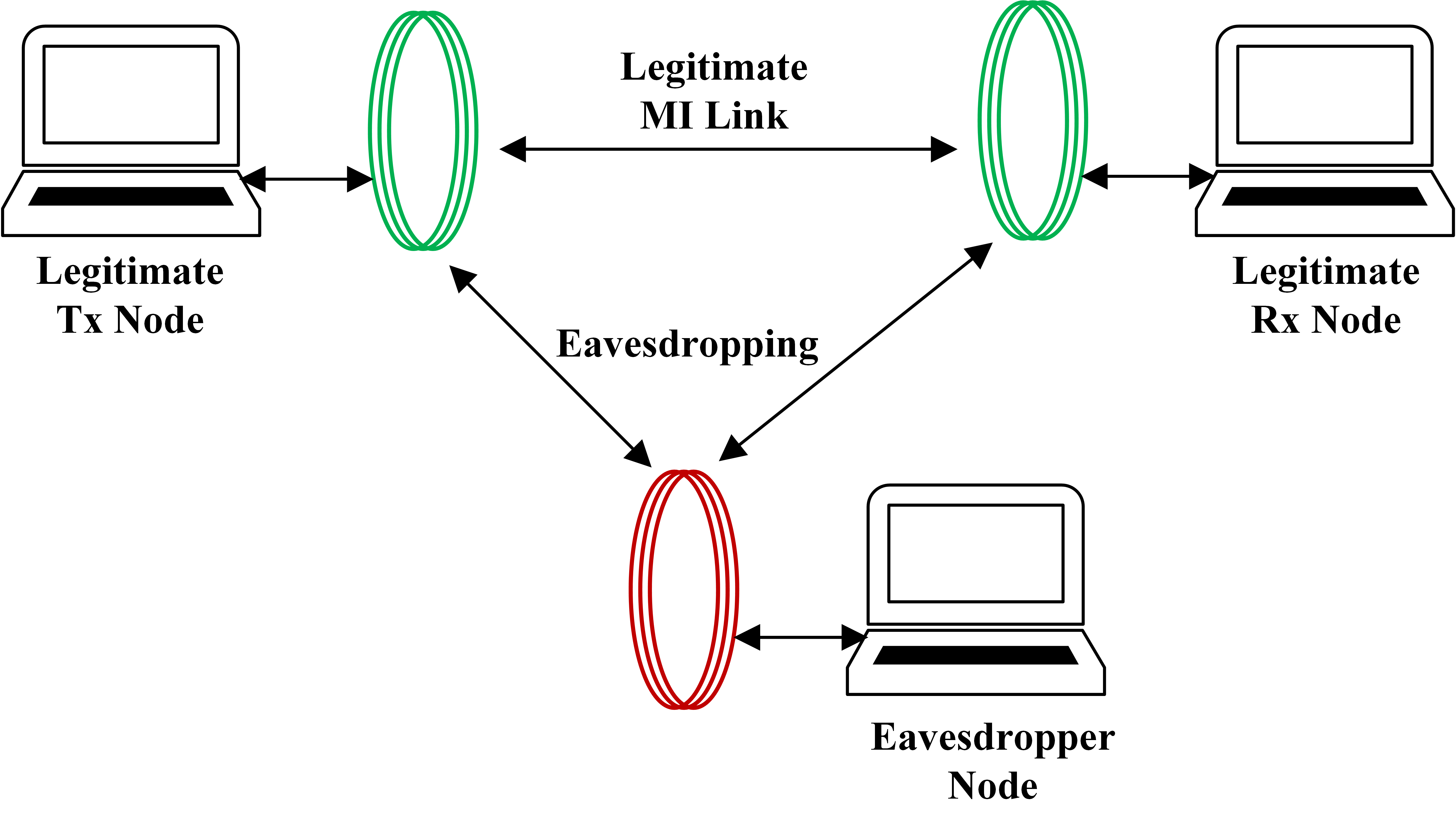}
    \caption{General illustration of an eavesdropping attack.}
    \label{fig:eavesattack}
\end{figure}

We consider MI communication between two legitimate nodes: Tx and Rx under the water with the presence of an eavesdropper node in closed proximity that aims to listen to the ongoing communication as shown in Fig. \ref{fig:eavesattack}. Since a time-varying sinusoidal signal is applied to the legitimate Tx node on the transmitter side, the legitimate Rx node and the eavesdropper node, therefore, receive the transmitted signal at the receiver side by coming into the vicinity of the generated magnetic field. Various configurations are used to evaluate the eavesdropping attack based on the eavesdropper node position and orientations (detail in Section \ref{sec:sim-results}).

\section{\textcolor{black}{Vulnerability Analysis against Eavesdropping Attacks}}

 \textcolor{black}{We first provide the details of the performance metrics used to investigate the vulnerability of underwater MI communication against eavesdropping attacks. Next, we present two types of quantitative vulnerability studies: one based on FEM simulations and the other based on real-world experiments. It is important to note that, due to the lack of a controlled setup for underwater MI communication experimentation, real-world experiments are conducted in an air medium. This approach is generally accepted, as MI communication is known to perform in a similar fashion across different environments \cite{li2019survey, muzzammil2020fundamentals, wei2020dynamic, domingo2012magnetic}.  However, slight variations in the quantitative values may still be observed due to differences in the conductivity of the mediums. These differences arise not only from the disparity in medium conductivity but also from minor mismatches in replicating the simulation model precisely in the real-world experimental setup. Such mismatches can result from uncertainties in hardware devices, including variations in component values such as capacitance and coil inductance, which may introduce additional deviations.
}
\label{sec:sim-results}
\subsection{Performance Metrics}
 We used two performance metrics: induced voltage and secret capacity to investigate the system's vulnerability against eavesdropping attacks. Below, we discuss each of them in detail.
\subsubsection{Induced Voltage}
When a signal is transmitted by magnetic induction, it induces a voltage in the receiving coil or conductor. The amplitude of this induced voltage is directly proportional to the strength of the magnetic field and the rate at which it is changing. Therefore, measuring the voltage in the receiver provides information on the strength of the signal that has been successfully transmitted. Additionally, the received power at the receiver coil is a function of the received voltage, as given in Eq. \ref{eq:SC}. Typically, the induced voltage can be expressed as given in Eq. \ref{eq:IV}
\subsubsection{Secrecy Capacity}
We consider secrecy capacity as a metric for this scenario where the secrecy capacity expression is given as:
\begin{align}
\label{eq:SC}
    \text{SC}=\log_2(1+\text{SNR}_\text{Rx})-\log_2(1+\text{SNR}_{\text{E}}),
\end{align}
where $\text{SNR}_{\text{Rx}}=(\frac{V_{\text{Rx}}^2}{R})/\sigma^2$ and $\text{SNR}_{E}=(\frac{V_{E}^2}{R})/{\sigma^2}$ are the received signal-to-noise ratio of legitimate Rx node and eavesdropper node respectively with $(\frac{V_{\text{Rx}}^2}{R})/(\frac{V_{E}^2}{R})$ the received power at the legitimate Rx/eavesdropper nodes, $\frac{V_{Rx}}{V_{E}}$ are the received voltages at legitimate Rx/eavesdropper, $R$ is the load resistance and $\sigma^2$ is the noise power. 

\begin{figure*}[h]
 \centering
    \subfigure[\label{subfig:config1Setup}]{\includegraphics[scale=0.65]{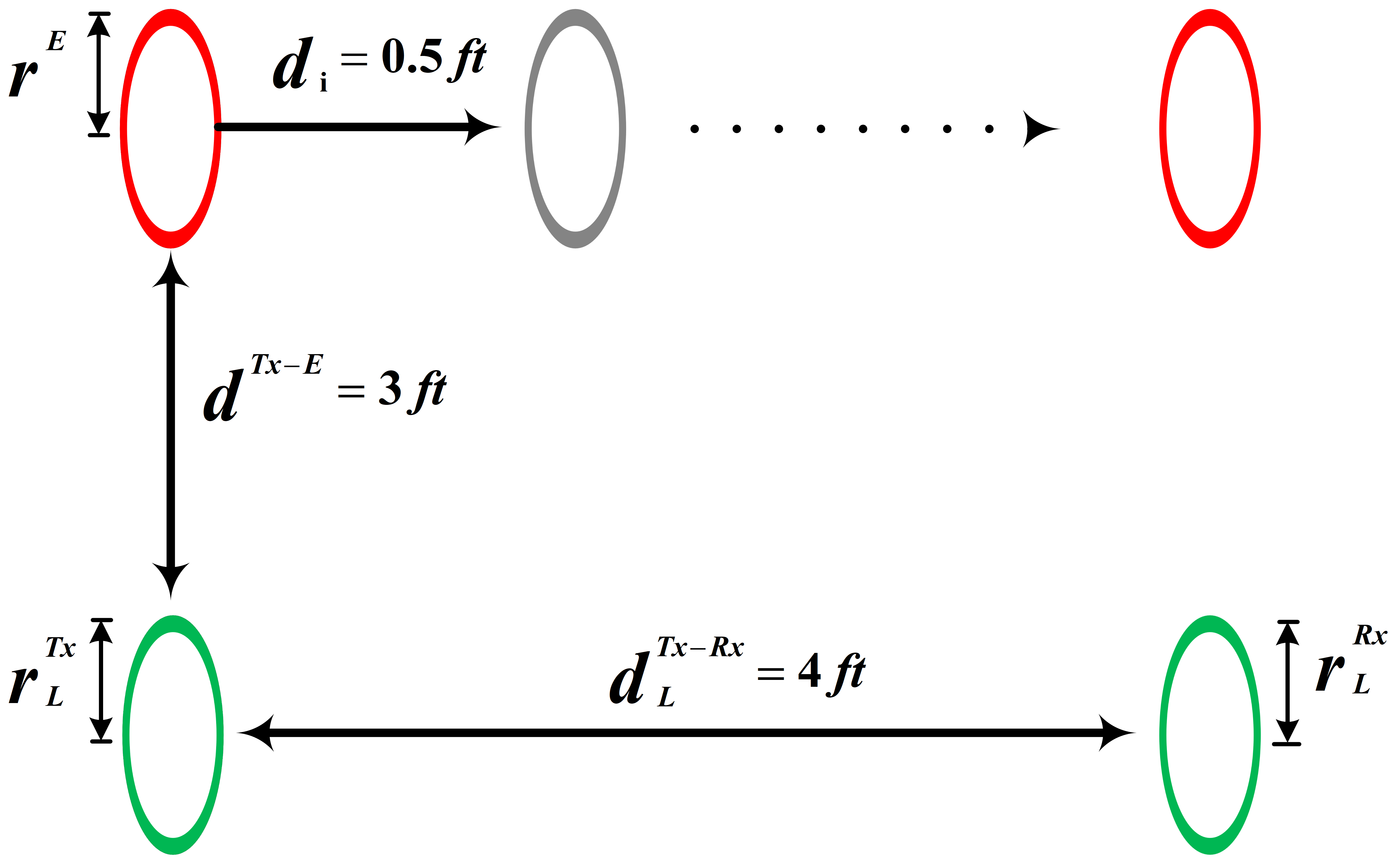}} 
  \subfigure[\label{subfig:config2Setup}]{\includegraphics[scale=0.65]{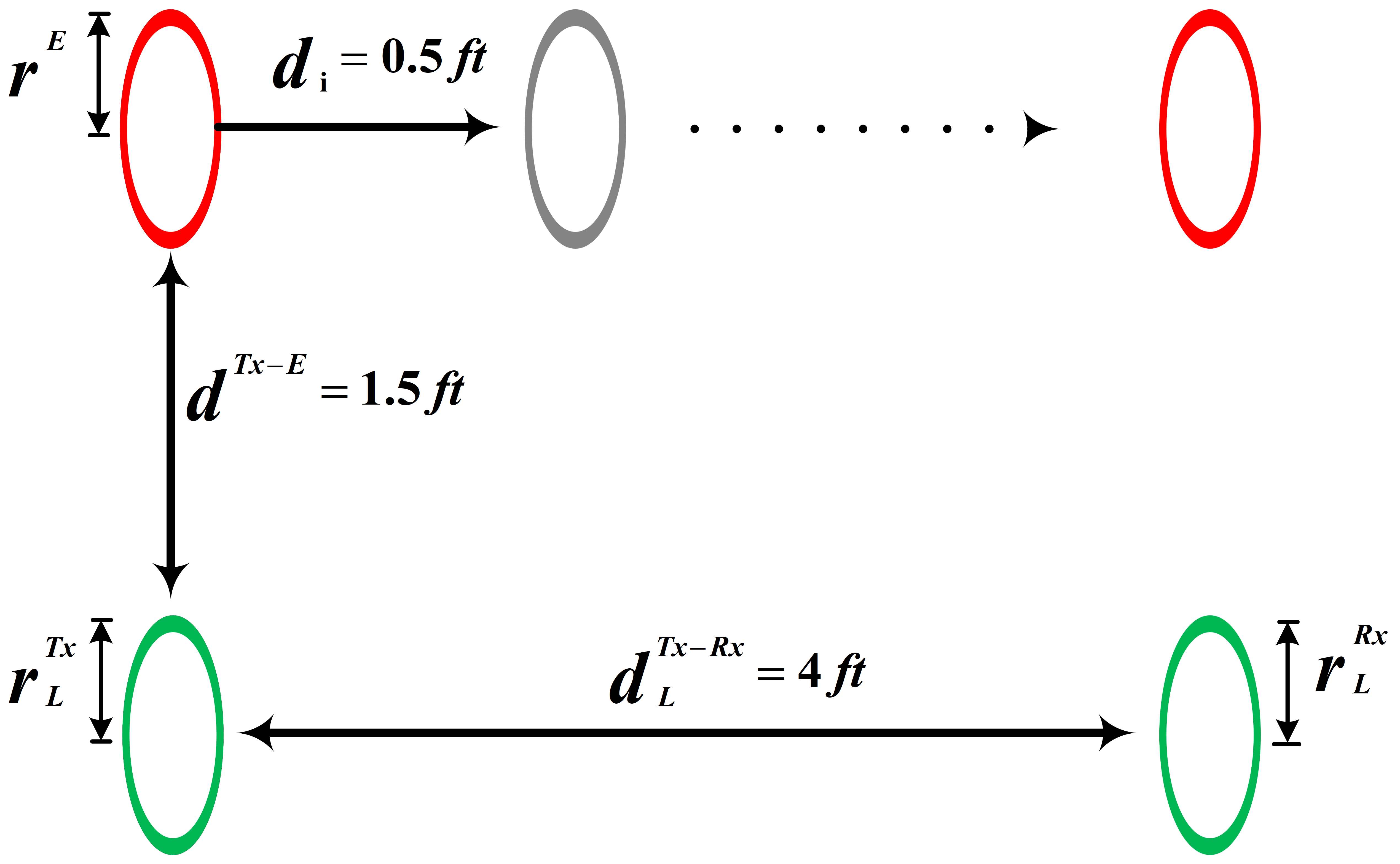}}
   \caption{Eavesdropper node position-based setup for both FEM simulation and lab experiments: (\textbf{a}) Configuration 1: When the eavesdropper node is placed far from the legitimate nodes and (\textbf{b}) Configuration 2: When the eavesdropper node is placed near to the legitimate nodes.}
 \label{fig:config1and2}
 \end{figure*}

 \begin{figure*}[h!]
 \centering
    \subfigure[\label{subfig:config1Vsim}]{\includegraphics[scale=0.60]{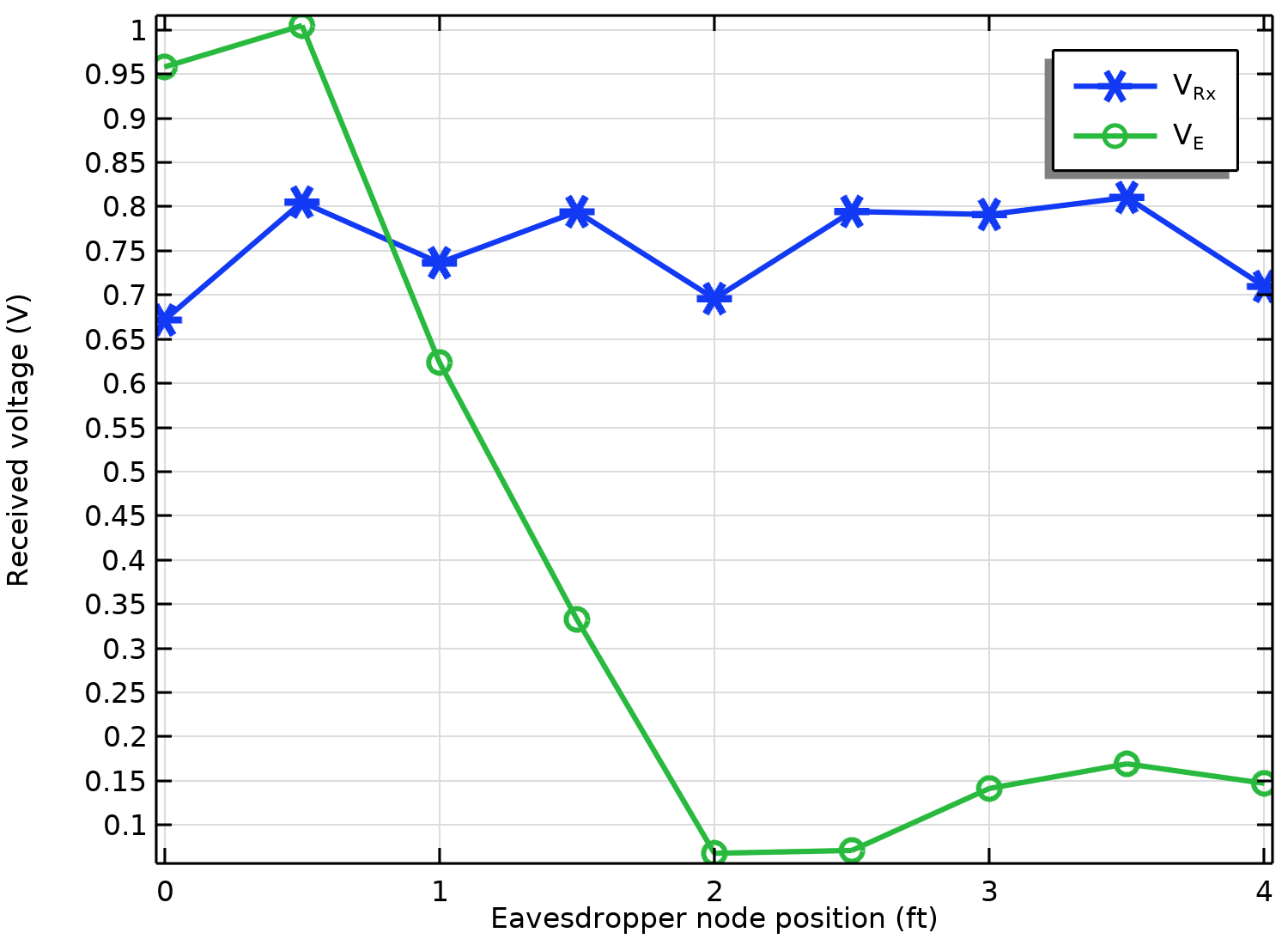}} 
  \subfigure[\label{subfig:config2Vsim}]{\includegraphics[scale=0.60]{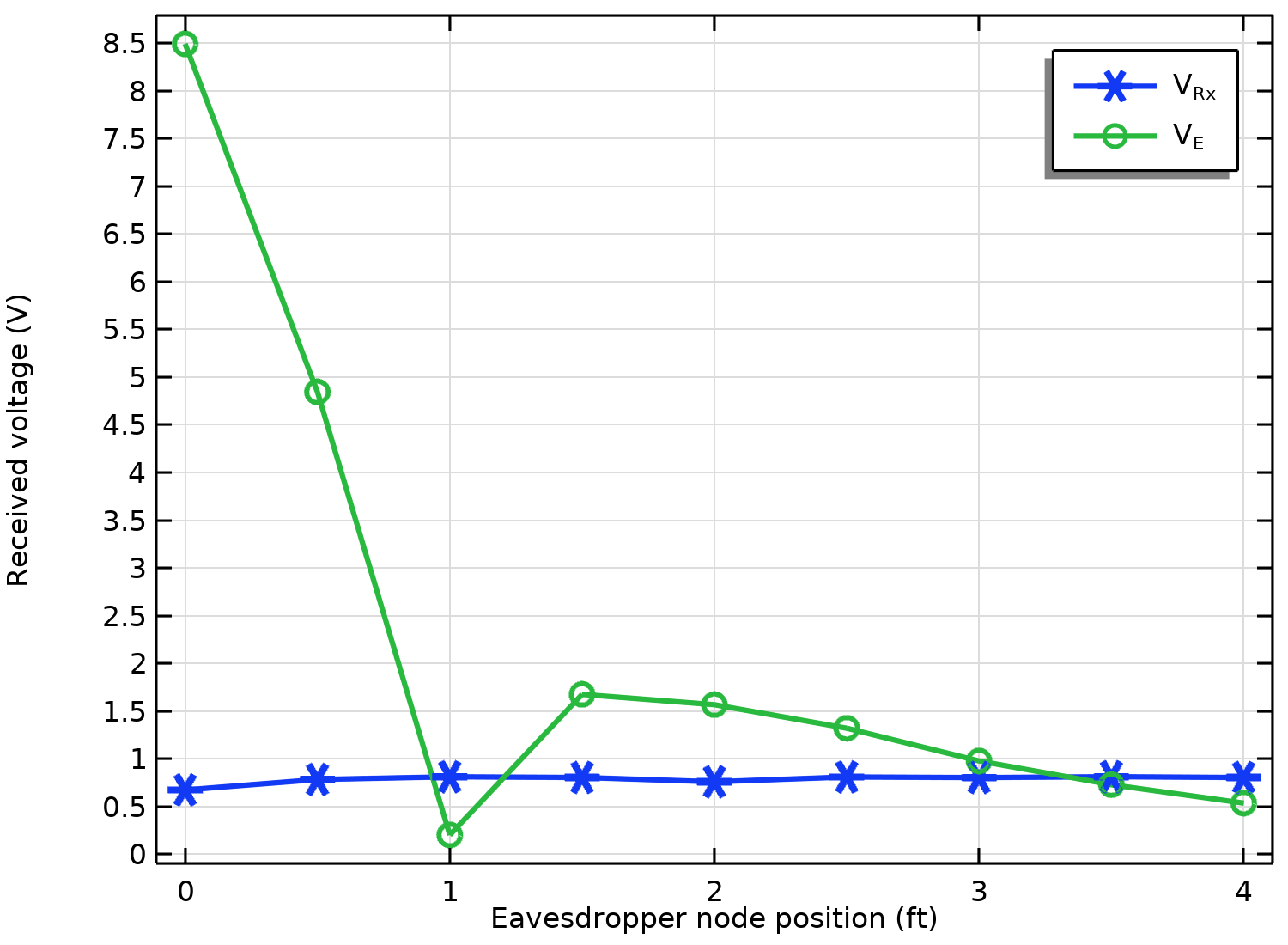}}
   \caption{\textcolor{black}{Simulation results based on eavesdropper node position with respect to legitimate Tx and Rx positions: (\textbf{a}) Received voltage vs. eavesdropper node position under Configuration 1 and (\textbf{b}) Received voltage vs. eavesdropper node position under configuration 2.}}
 \label{fig:simEavesPosition}
 \end{figure*}

\subsection{\textcolor{black}{FEM-based Simulation}}
In our simulation experiments, we use FEM simulation to develop the MI communication security model discussed in Section~\ref{sec:eaves}. 
A sinusoidal voltage signal of $10V$ is applied to Tx. The radius of the coil and the number of turns in the legitimate Tx, legitimate Rx, and eavesdropper nodes remain the same, which is $12.7cm$ and $30$, respectively. The capacitor is attached in series with the Tx coil but parallel to the legitimate Rx node and the eavesdropper node to resonate at about $100kHz$ frequency. The capacitance values of the capacitor and coil inductance are listed in Table~\ref{tab:sim-tab}. Finally, the conductivity of water is fixed to $0.01S/m$. \textcolor{black}{The parameter values utilized in this study are provided in Table. \ref{tab:sim-tab}, which are the same in overall simulations unless explicitly mentioned.} 
%

\begin{table}[h]
    \centering
   \caption{\textcolor{black}{Parameter settings used in simulations setup.}}
    \label{tab:sim-tab}
    \begin{tabular}{|c|c|c|c|}
    \hline
         & Tx node & Rx node & Eavesdropper node\\
     \hline  
    Coil radius & 12.7 $cm$ & 12.7 $cm$ & 12.7 $cm$ \\
    \hline
    Number of turns & 30 & 30 & 30 \\
    \hline
    Capacitance    & 7.681 $nF$  &  7.681$nF$ & 7.681$nF$ \\
     \hline    
    Inductance   & 329.75 $\mu H$ & 329.75 $\mu H$ & 329.75 $\mu H$ \\
    \hline   
    Voltage applied & 10 V & - & - \\
    \hline
    Resonance frequency & 100 $KHz$ & 100 $KHz$ & 100 $KHz$ \\
    \hline
     Load Resister & - & 50 $\Omega$ & 50 $\Omega$ \\
    \hline
    \end{tabular}
 
\end{table}

In the following section, we evaluate the eavesdropping attack based on the position of the eavesdropper node, the orientation of the eavesdropper node, and the secrecy capacity. 

 \begin{figure*}[]
 \centering
    \subfigure[\label{subfig:config3C31_setup}]{\includegraphics[scale=0.67]{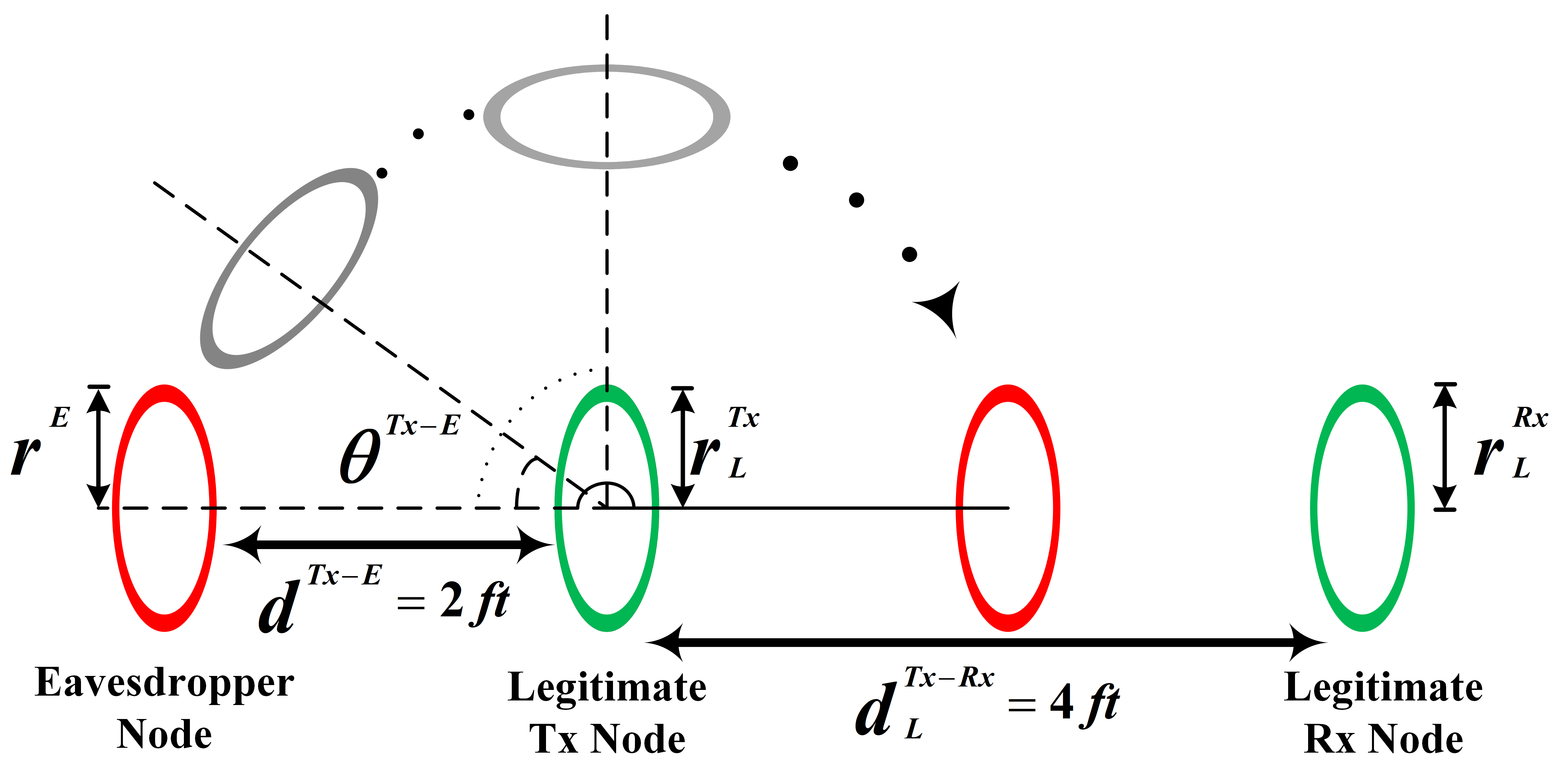}}
    \subfigure[\label{subfig:config3C32_setup}]{\includegraphics[scale=0.67]{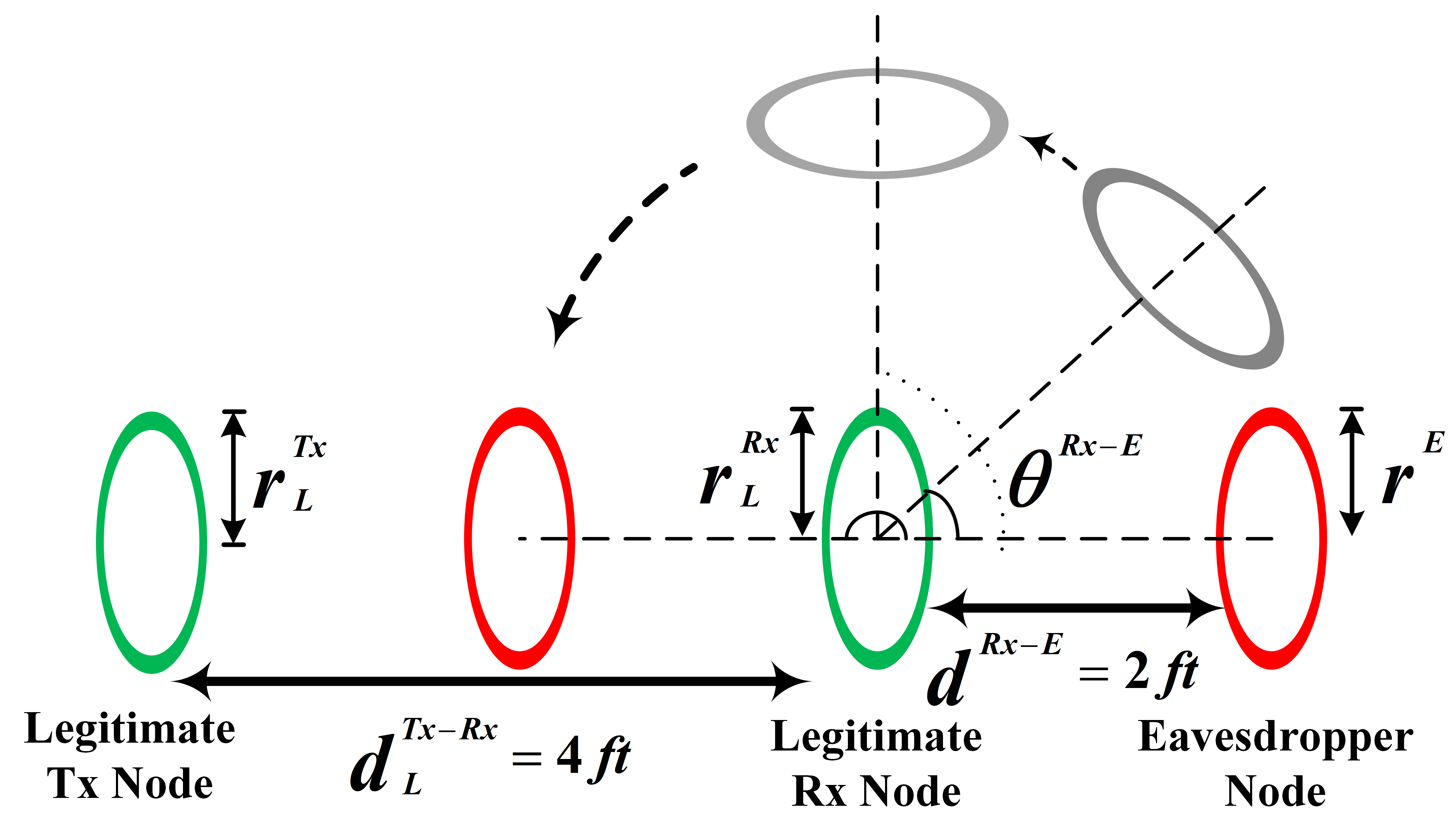}} 
    \subfigure[\label{subfig:config3C33_setup}]{\includegraphics[scale=0.77]{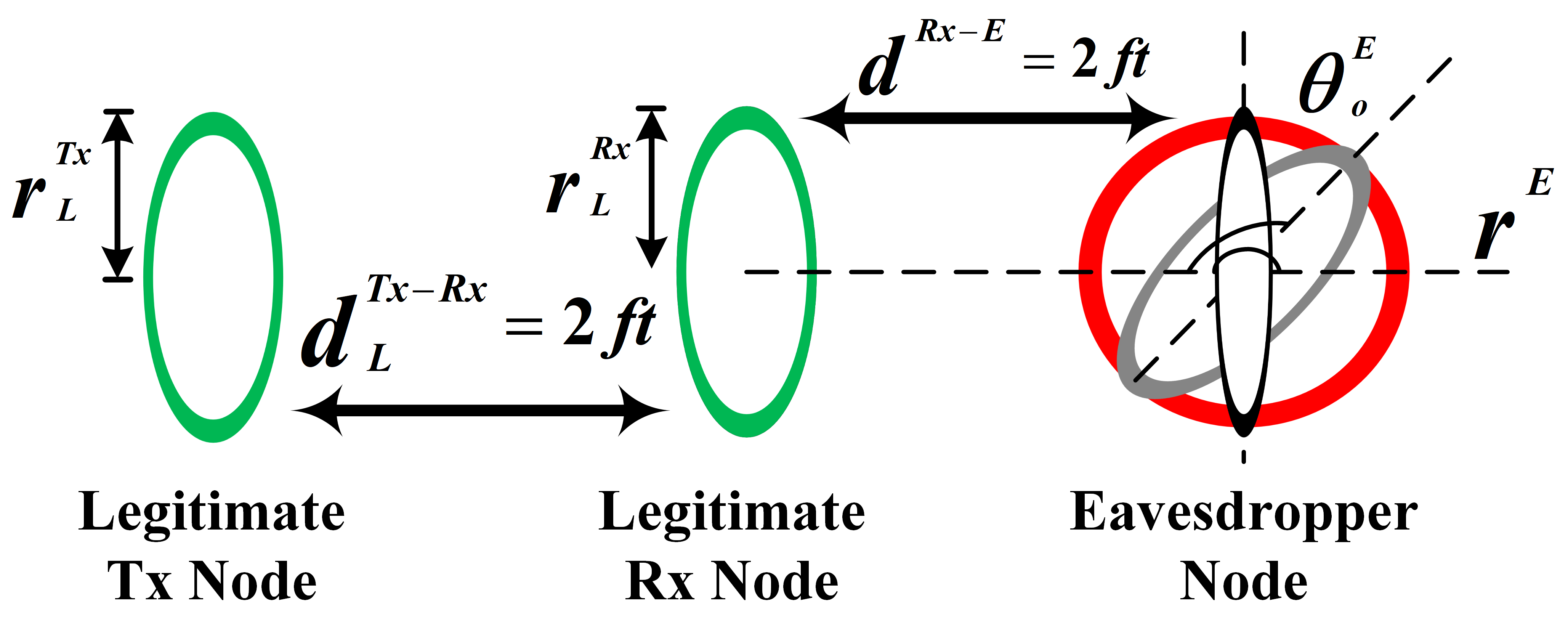}} 
 \caption{Eavesdropper node orientation changes (\textbf{a}) \textcolor{black}{Configuration 3}: eavesdropper changes its position w.r.t. legitimate Tx node, (\textbf{b}) \textcolor{black}{Configuration 4}: eavesdropper changes its position w.r.t. legitimate Rx node, and (\textbf{c}) \textcolor{black}{Configuration 5: eavesdropper changes its position w.r.t. its own origin.}}
 \label{fig:config3_setup}
 \end{figure*}

\subsubsection{Eavesdropper node position}
\label{sec:sim-Eposition}
In this section, the eavesdropping attack is studied based on the position/placement of the eavesdropper node with respect to legitimate nodes. The position of an eavesdropper node is crucial, as the closer the eavesdropper node is to the legitimate transmitted magnetic field, the more information can be eavesdropped. To study the impact of the position of the eavesdropper node, we consider two different configurations based on how far the eavsdropper node from the legitimate Tx and Rx nodes as shown in Fig.~\ref{fig:config1and2}. FEM simulation models are developed for two different configurations as depicted in Fig.~\ref{fig:config1and2}. The detail of the two configurations are below: 

\begin{enumerate}
    \item \textit{Configuration 1:} In this configuration, the eavesdropper is placed far from the legitimate nodes. Initially, the legitimate Tx and Rx nodes are fixed at points (0,0,0)ft and (4,0,0)ft (i.e., $d_{L}^{Tx-Rx}=4ft$), respectively. While, the eavesdropper node is initially placed at (0,3,0)ft from the Tx node (i.e., $d^{Tx-E}=3ft$), and then moves away at an interval of 0.5ft till (4,3,0)ft as depicted in Fig.~\ref{subfig:config1Setup}.
    \textcolor{black}{\item \textit{Configuration 2:} In this configuration, the eavesdropper node is placed closer to the legitimate Tx and Rx nodes. The transmitter node is kept fixed at a point of (0,0,0)ft while a legitimate Rx node is fixed at (4,0,0)ft (i.e., $d_{L}^{Tx-Rx}=4ft$). The eavesdropper node is initially placed at (0,1.5,0)ft from the Tx node (i.e., $d^{Tx-E}=1.5ft$), and then moves away at an interval of 0.5ft till (4,1.5,0)ft as depicted in Fig.~\ref{subfig:config2Setup}.}
\end{enumerate}

Fig.~\ref{fig:simEavesPosition} shows the received voltages vs. eavesdropper node positions under both configuration 1 and 2. In Fig.~\ref{subfig:config1Vsim}, received voltages of the legitimate Rx node (denoted by $V_{Rx}$) and the eavesdropper node (denoted by $V_E$) are shown against different positions of the eavesdropper node under configuration 1. It can be seen that the legitimate Rx node experiences changes in its received voltage against different positions of the eavesdropper node. It can also be seen that the eavesdropper node itself receives a different voltage at different positions. Maximum voltage received in the eavesdropper node at positions (0,3,0)ft and (0.5,3,0)ft, while minimum voltage received in the eavesdropper node at positions (2,3,0)ft and (2.5,3,0)ft. This means that the position of the eavesdropper node influences the eavesdropping process on the ongoing legitimate MI communication. That is, in some positions, the eavesdropper node may receive maximum information, while in others, it may receive less information due to the low-voltage signal it receives. Furthermore, in this configuration, it is quite possible that the legitimate nodes can sense malicious activities due to the change in the received voltage in the legitimate Rx node and, therefore, can apply countermeasure techniques.

Similarly, Fig.~\ref{subfig:config2Vsim} shows the voltage received in the legitimate Rx node and eavesdropper node versus different positions of the eavesdropper node based on the configuration 2 setup. The legitimate Rx node exhibits little to no change in the received voltage, as shown in Fig.~\ref{subfig:config2Vsim}. The maximum received voltage in the eavesdropper node in this configuration can be seen in the position (0,1.5,0) feet. This is expected because the eavesdropper node is now half a distance closer to the legitimate Tx node and, therefore, receives a higher voltage than Fig. \ref{subfig:config1Vsim} in configuration 1. However, the eavesdropper node received negligible voltage at position (1,1.5,0)ft and may not receive any information at this position from ongoing MI communication.

To conclude, in both configurations 1 and 2, the eavesdropper node can receive information from the ongoing legitimate MI communication. The amount of information the eavesdropper node can get depends on its position with respect to the legitimate Tx node. Furthermore, due to the slight changes in the received voltage of the legitimate Rx node, Rx can detect malicious activity and apply countermeasures accordingly. Therefore, receiver sensitivity and receiver processing capabilities can play an important role.

\subsubsection{Eavesdropper node orientation}
\label{sec:sim-Eorientation}
Since the orientation of coils can be changed in the underwater environment due to water movements and tides, it is important to study the impact of the coil orientations on the eavesdropping attack. In addition to the natural changes in the orientation of the coils, the eavesdropper may be smart enough to change its orientation to receive higher voltage and, therefore, eavesdrop more information. 

To study the impact of the orientation of the eavesdropper node with respect to legitimate nodes, we consider three different configurations, as shown in Fig. \ref{fig:config3_setup}. In all three configurations, the legitimate Tx and Rx nodes are kept fixed at a distance of 4ft, with positions (0,0,0)ft and (4,0,0)ft, respectively, unless explicitly mentioned. As legitimate Tx and Rx coils face each other, the angle $\theta_{L}^{Tx-Rx}=0^\circ$ receives the maximum signal strength. However, the position and angle of the eavesdropper node change in a rotational fashion with respect to the legitimate Tx node, the legitimate Rx node, and its own origin, which we model in the following configurations. 

\subsubsection{\textcolor{black}{Configuration 3}}
\label{sec:OConfig1}
In this configuration, the eavesdropper node is initially placed 2ft away from the legitimate Tx node at (-2,0,0)ft (i.e., $d^{Tx-E}=2ft$), where the angle between them is initially $\theta^{Tx-E}=0^\circ$. The angle is then rotated at an interval of \textcolor{black}{$\theta=30^\circ$} from $0^\circ$ to $180^\circ$ around the legitimate Tx node as shown in Fig.\ref{subfig:config3C31_setup}.

 \begin{figure*}[]
 \centering
    \subfigure[\label{subfig:config3C11_MFD}]{\includegraphics[width=0.30\textwidth]{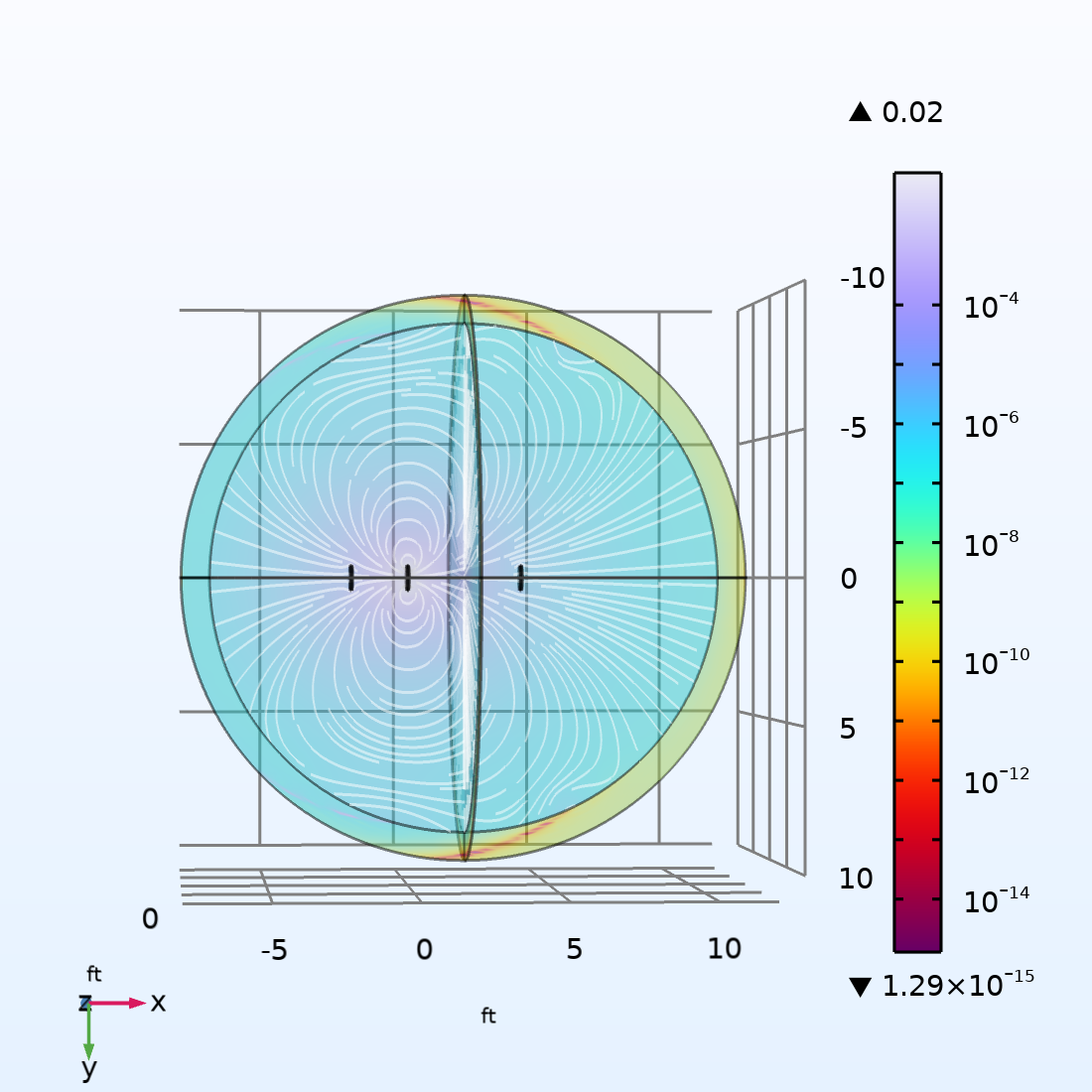}}
    \subfigure[\label{subfig:config3C12_MFD}]{\includegraphics[width=0.30\textwidth]{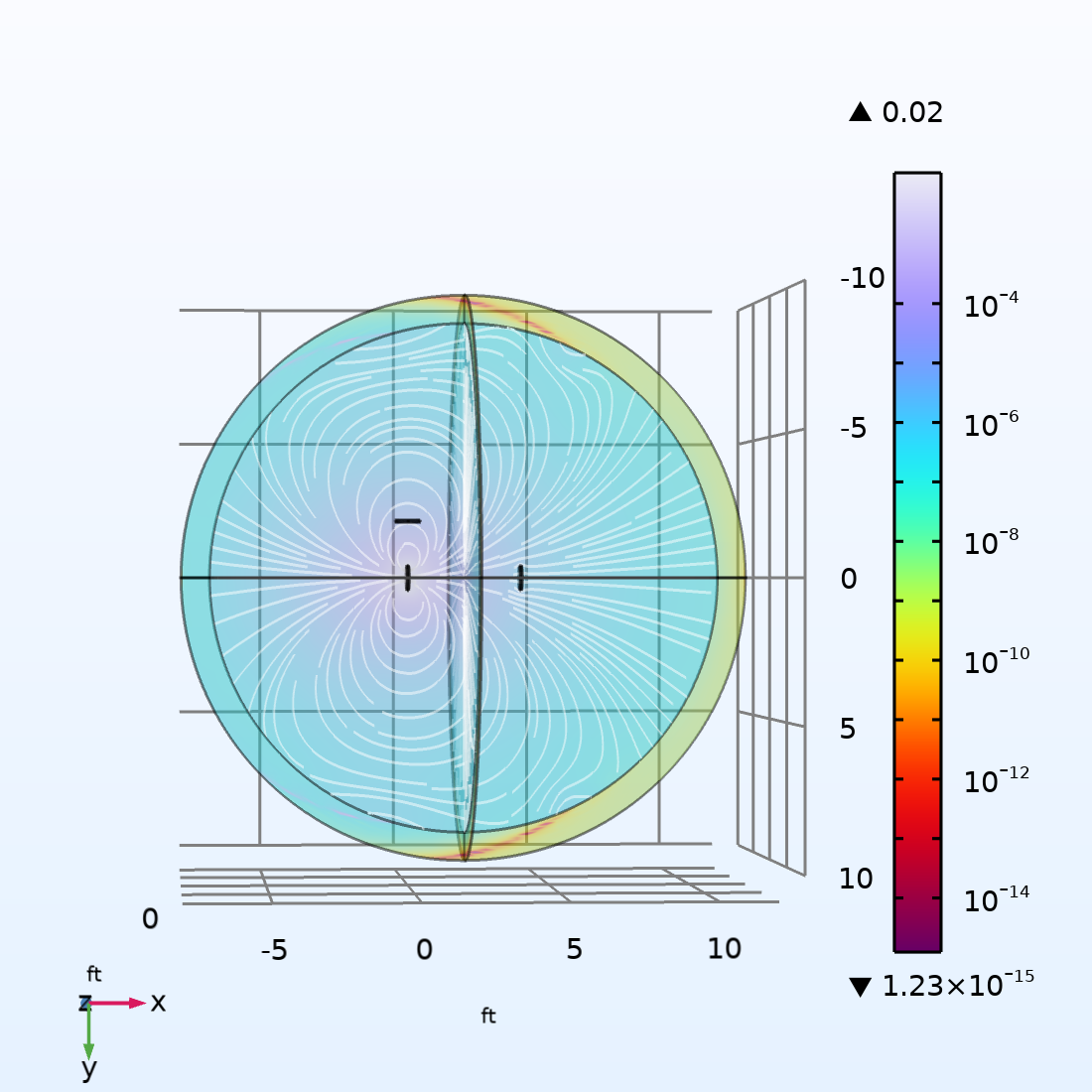}} 
    \subfigure[\label{subfig:config3C13_MFD}]{\includegraphics[width=0.30\textwidth]{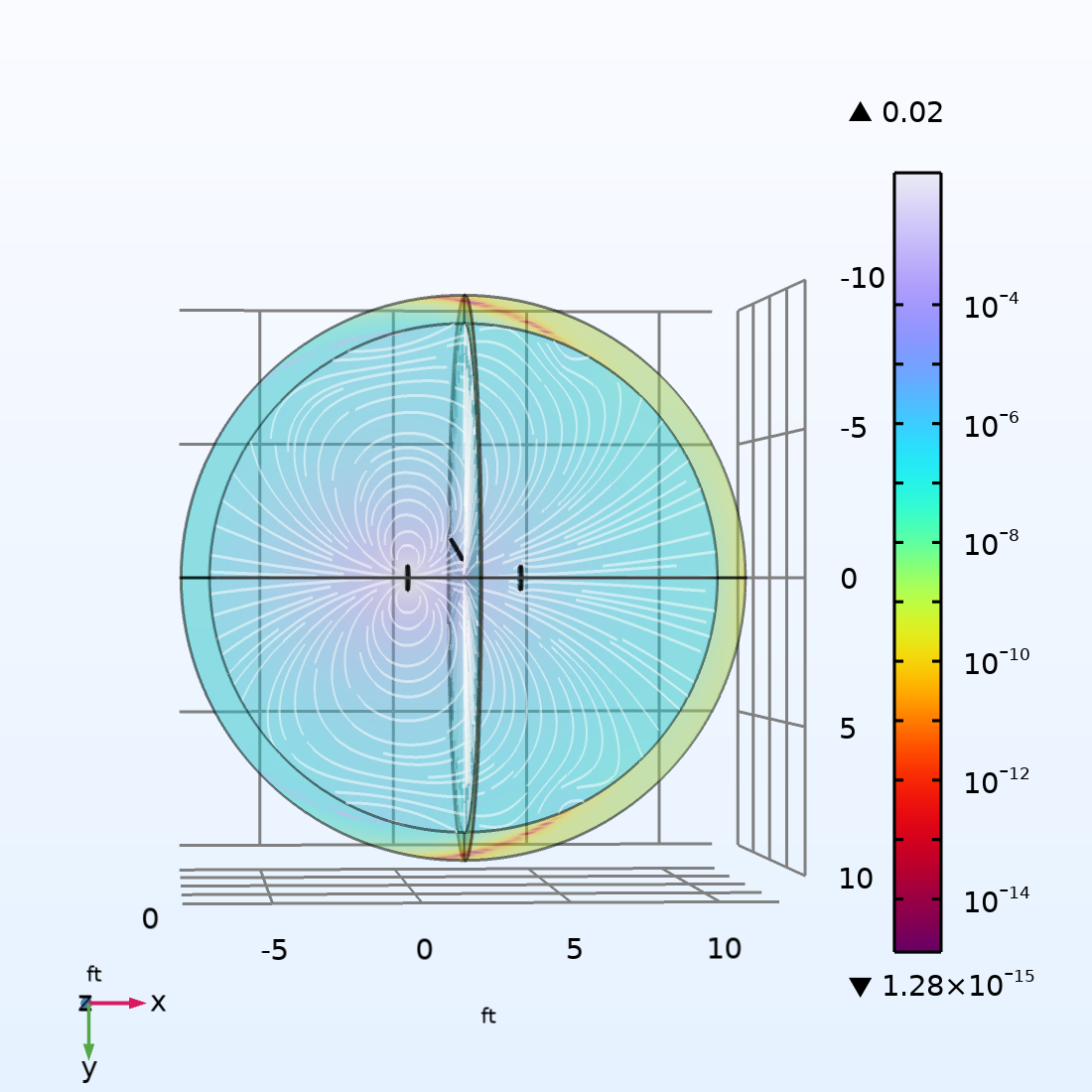}} 
 \caption{Magnetic flux density norm in $T$ with respect to different eavesdropper node angle in the case of configuration 3 when: (\textbf{a}) $\theta^{Tx-E}=0^{\circ}$, (\textbf{b}) $\theta^{Tx-E}=90^{\circ}$, and \textcolor{black}{(\textbf{c}) $\theta^{Tx-E}=150^{\circ}$.}}
 \label{fig:config3C1_MFD}
 \end{figure*}

 \begin{figure}[]
    \centering
    \includegraphics[scale=0.60]{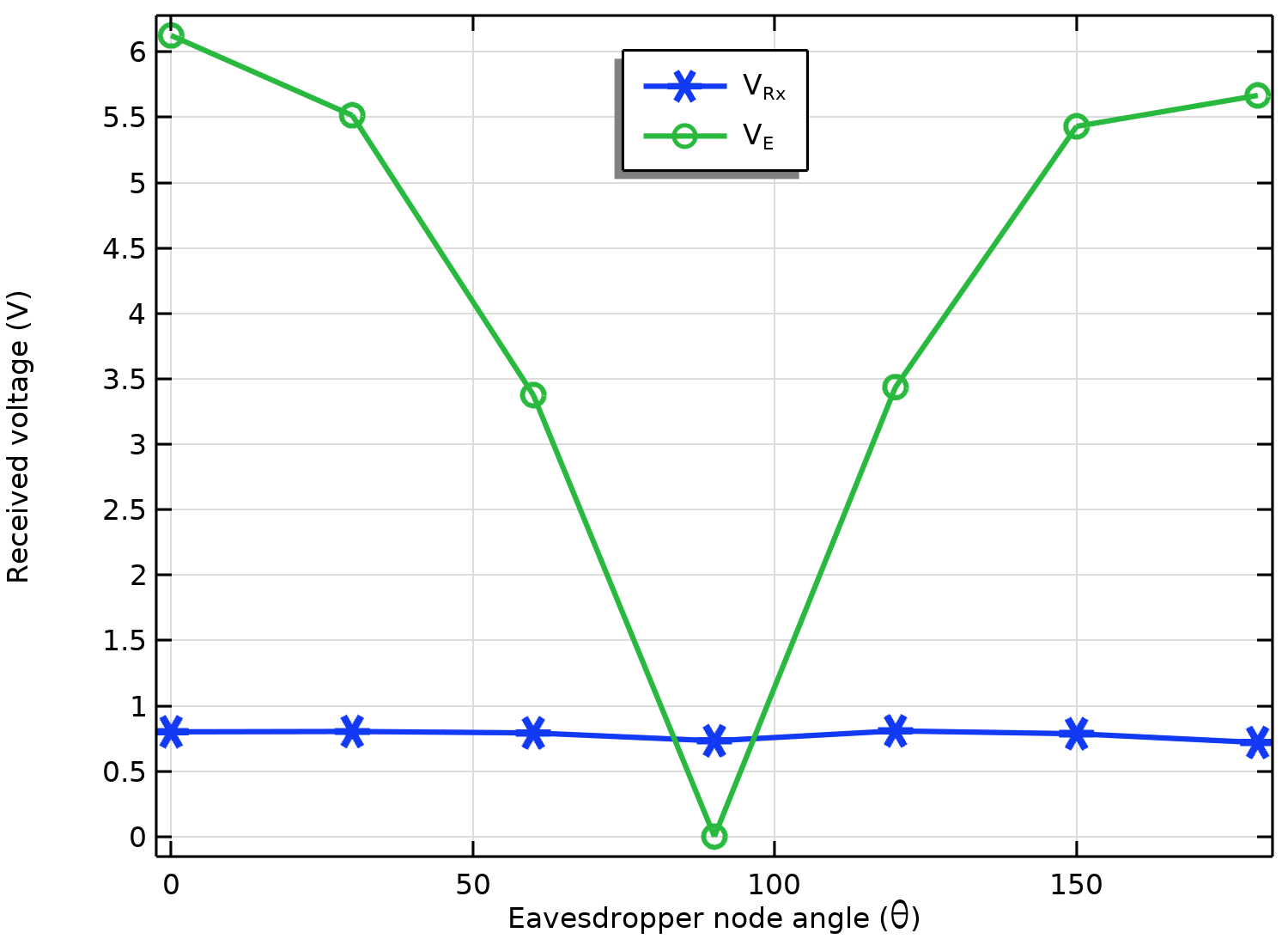}
    \caption{\textcolor{black}{Received voltage in legitimate Rx node and eavesdropper node vs. different eavesdropper node orientation by changing its angle in a rotational fashion w.r.t. legitimate Tx node position - configuration 3.}}
    \label{fig:config3C1_anglewrtTxnode}
\end{figure}

In Fig.~\ref{fig:config3C1_MFD}, the magnetic flux density is shown based on \textcolor{black}{configuration 3}, where the eavesdropper node had a different orientation from the legitimate Tx node. Although there is no significant change in the overall magnetic flux density, the magnetic flux (which is $\phi = BAcos\theta$) would be different in each case of Fig. \ref{subfig:config3C11_MFD}, \ref{subfig:config3C12_MFD}, and \ref{subfig:config3C13_MFD} because the orientation of the eavesdropper node with respect to the magnetic field lines and magnetic flux density at that position plays a significant role. For the case of Fig. \ref{subfig:config3C11_MFD} and \ref{subfig:config3C13_MFD}, the magnetic flux is stronger because the magnetic flux density is relatively high and the magnetic field lines coming toward the eavesdropper node area are with some angle, i.e. $\theta^{Tx-E} \neq 0$. As shown in Fig. \ref{subfig:config3C12_MFD}, magnetic flux approaches zero because the magnetic field lines and the coil orientation are parallel to each other, i.e., $\theta^{Tx-E} = 90^{\circ}$.

Unlike the magnetic flux density results, Fig.~\ref{fig:config3C1_anglewrtTxnode} illustrates how much voltage can be received by the eavesdropper node due to a different orientation with respect to the legitimate Tx node. Furthermore, the impact of the orientation of the eavesdropper node on the received voltage of the legitimate Rx node must be identified. From  Fig.~\ref{fig:config3C1_anglewrtTxnode}, maximum voltage received can be seen at the eavesdropper angles $\theta^{Tx-E}=0^\circ$, \textcolor{black}{$\theta^{Tx-E}=30^\circ$, and $\theta^{Tx-E}=150^\circ$}, which shows that the eavesdropper node is approximately face-to-face to the Tx node, and hence will receive maximum information. When the angle of the eavesdropper node to the Tx node becomes orthogonal, that is, $\theta^{Tx-E}=90^\circ$, no voltage is received at the eavesdropper node, and therefore, no information can be retrieved from ongoing legitimate MI communication. On the other hand, minimal to no change can be seen in the voltage received from the legitimate Rx node, except in $\theta^{Tx-E}=90^\circ$ and $\theta^{Tx-E}=180^\circ$. That means a decoupling occurs to some extent between the legitimate nodes due to the existence of the eavesdropper node at these angles. Consequently, legitimate nodes can detect malicious activity resulting from this change in the received voltage.

\subsubsection{\textcolor{black}{Configuration 4}}
\label{sec:OConfig2}
In this configuration, the eavesdropper node is initially placed 2ft away from the legitimate Rx node at (6,0,0)ft (i.e., $d^{Tx-E}=6ft$), where, initially, the angle between them $\theta^{Rx-E}=0^\circ$. We note that the respective angle between legitimate Tx and the eavesdropper node is also initially $\theta^{Tx-E}=0^\circ$. Then the angle of the eavesdropper node is rotated in an interval of \textcolor{black}{$\theta=30^\circ$}  from $0^\circ$ to $180^\circ$ around the legitimate Rx node as shown in Fig.~\ref{subfig:config3C32_setup}.

\begin{figure*}[h!]
 \centering
    \subfigure[\label{subfig:config3C21_MFD}]{\includegraphics[width=0.30\textwidth]{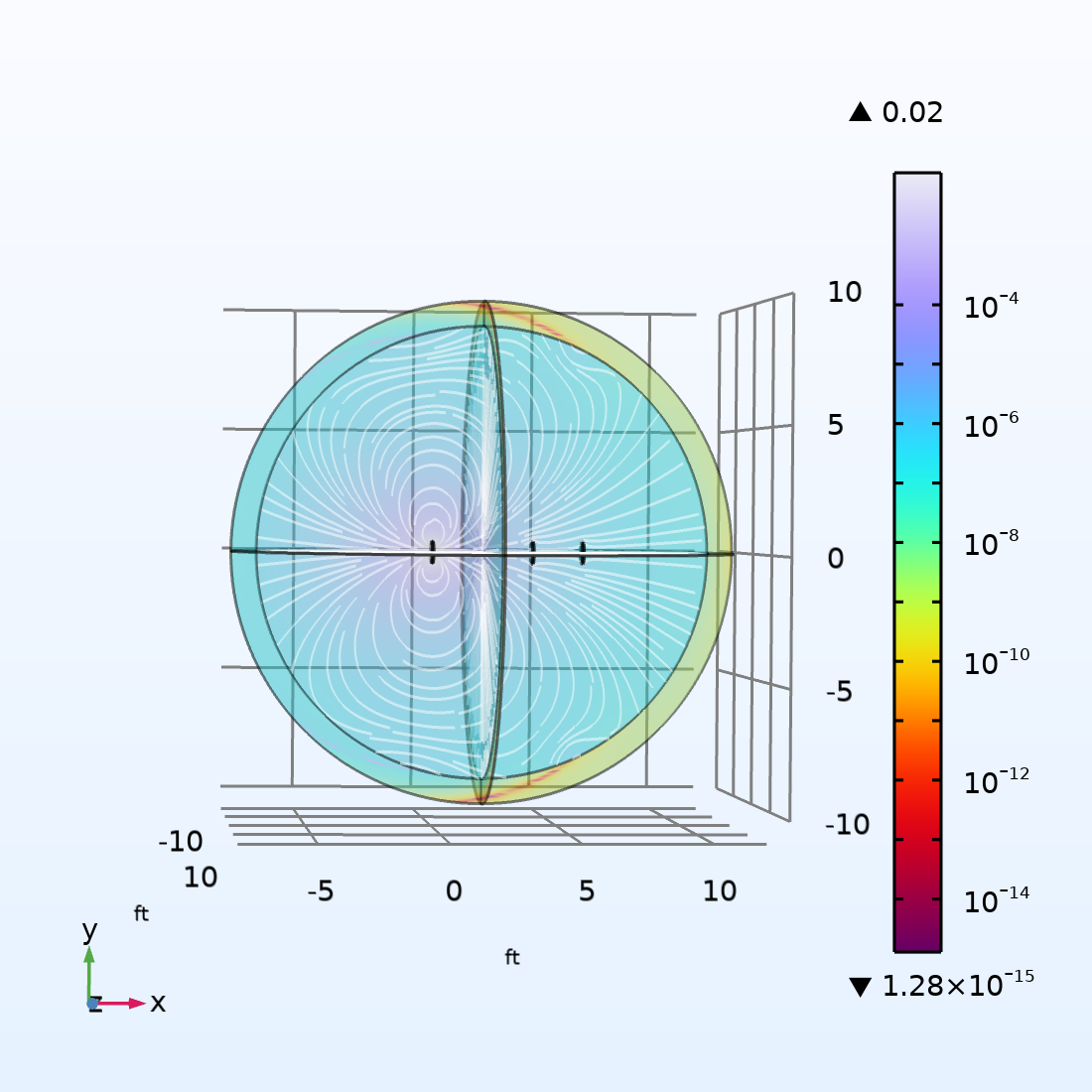}}
    \subfigure[\label{subfig:config3C22_MFD}]{\includegraphics[width=0.30\textwidth]{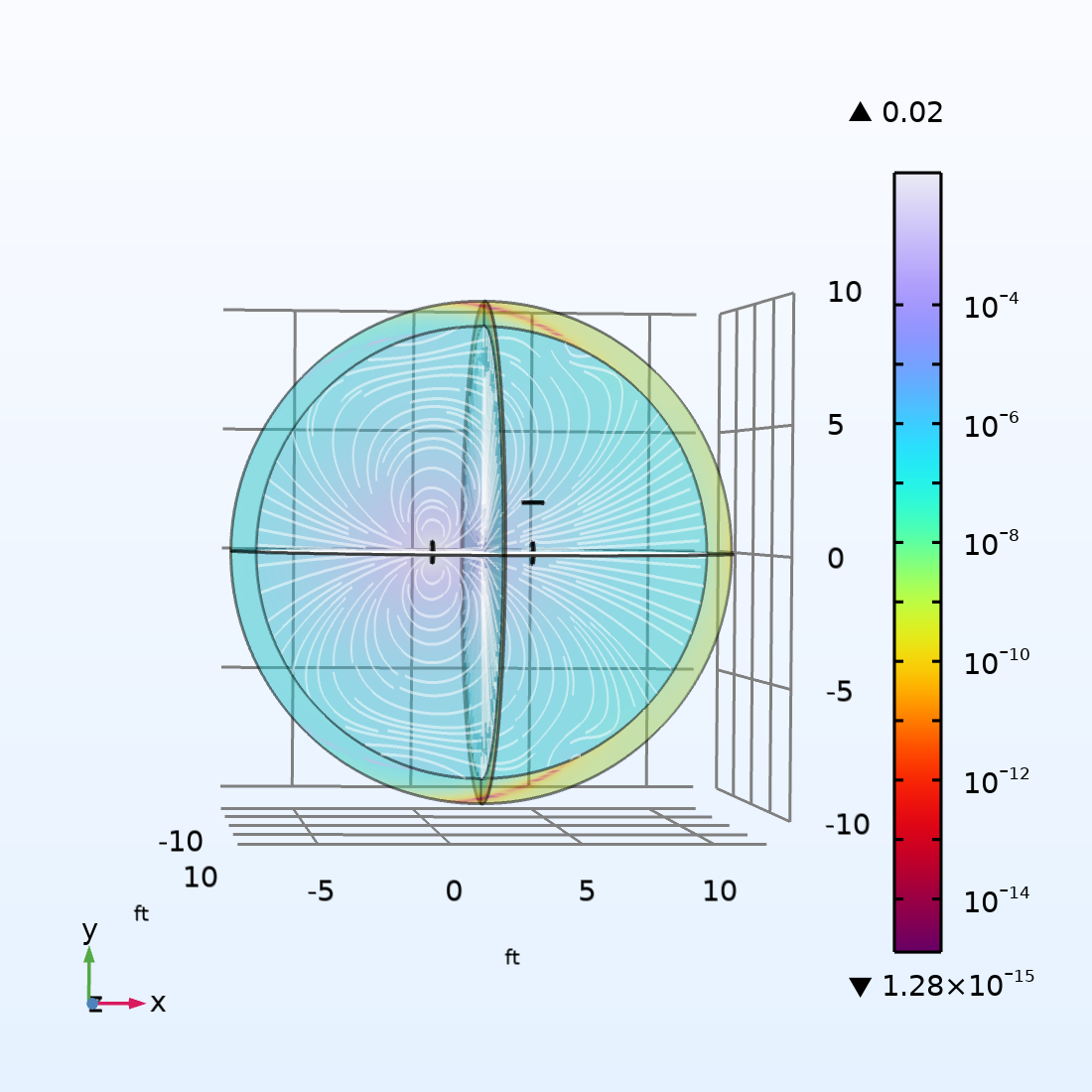}} 
    \subfigure[\label{subfig:config3C23_MFD}]{\includegraphics[width=0.30\textwidth]{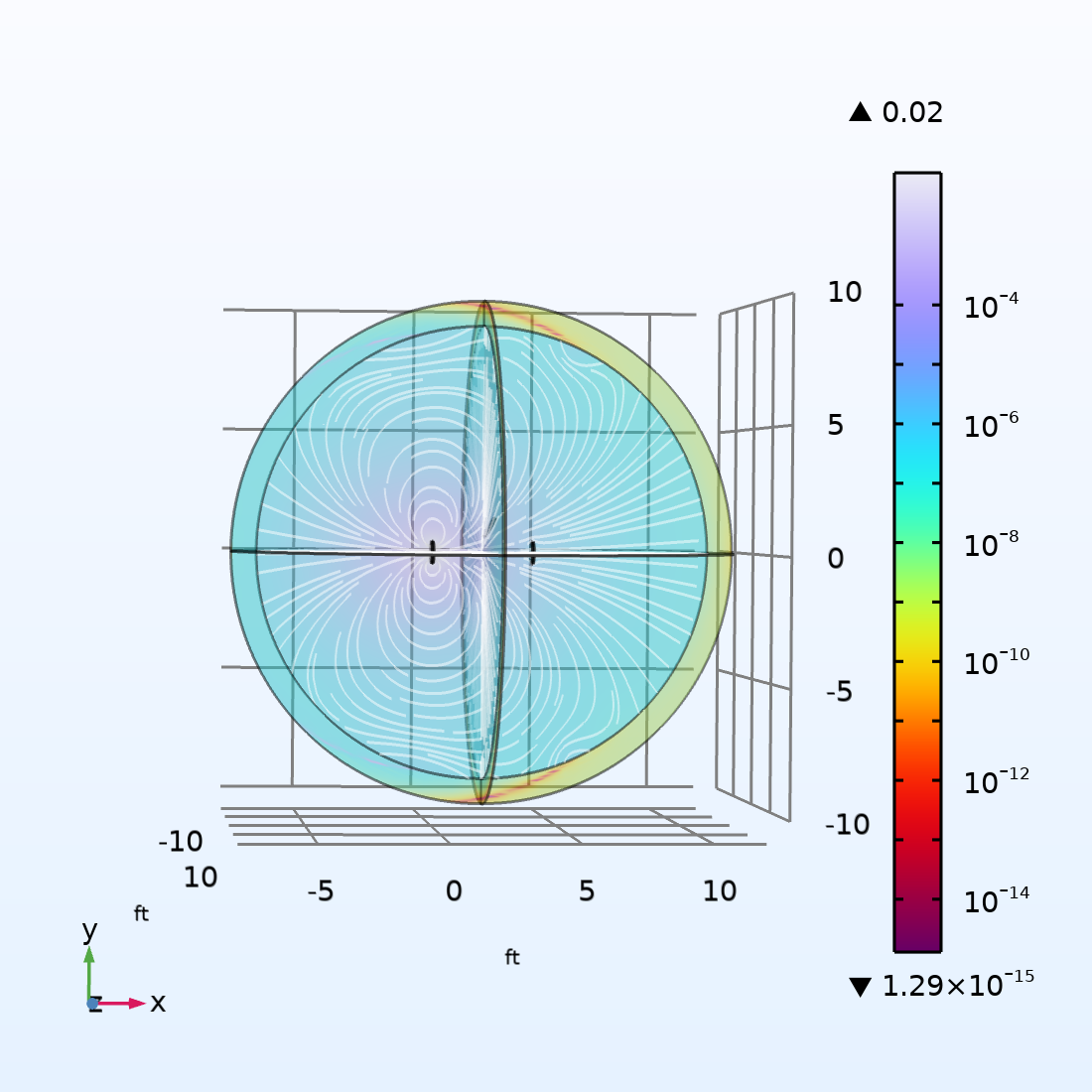}} 
 \caption{Magnetic flux density norm in $T$ with respect to different eavesdropper node angle in the case of configuration 4 when: (\textbf{a}) $\theta^{Rx-E}=0^{\circ}$, (\textbf{b}) $\theta^{Rx-E}=90^{\circ}$, and (\textbf{c}) \textcolor{black}{$\theta^{Rx-E}=150^{\circ}$.}}
 \label{fig:config3C2_MFD}
 \end{figure*}
 
  \begin{figure}[h!]
    \centering
    \includegraphics[scale=0.60]{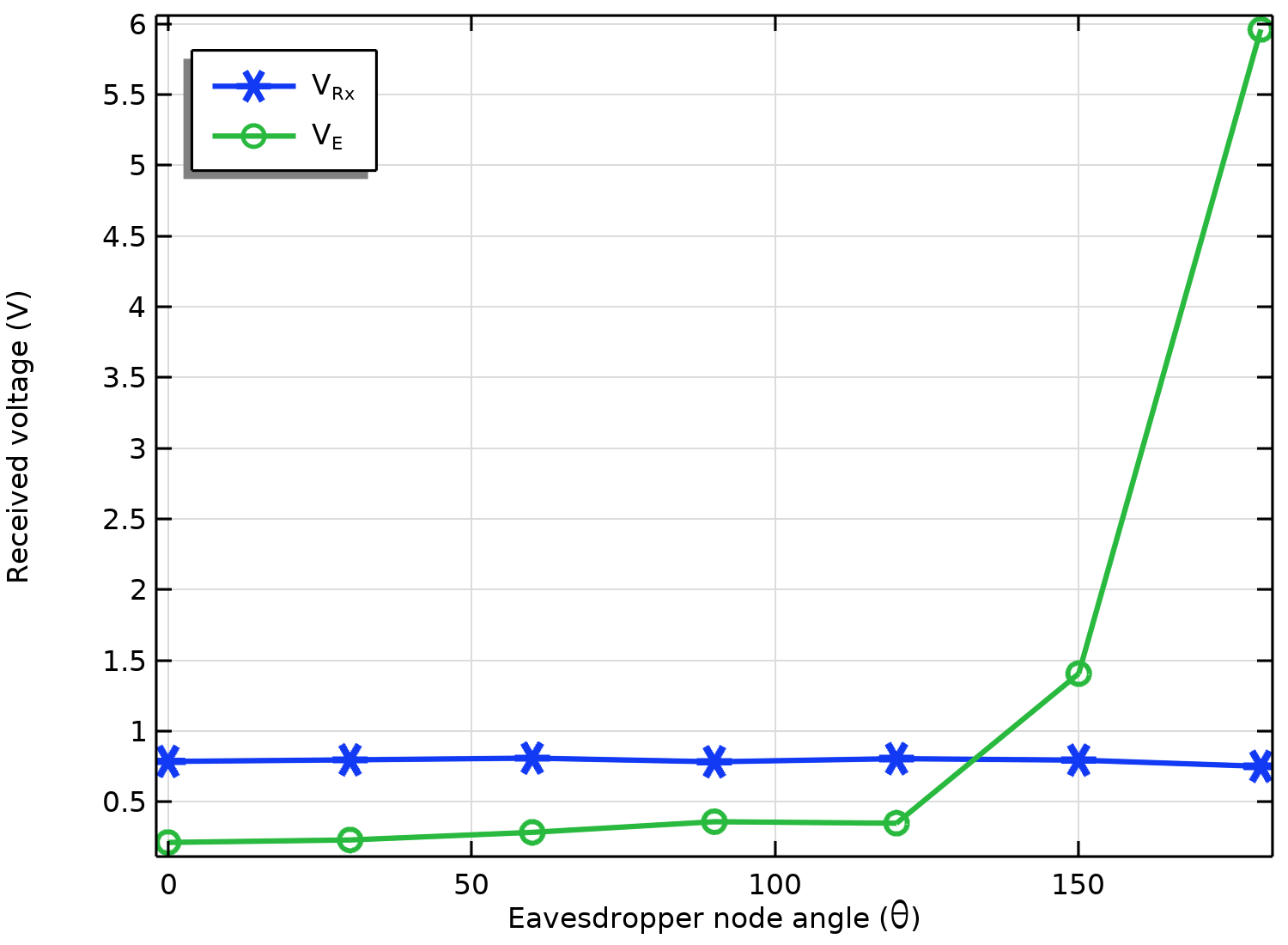}
    \caption{\textcolor{black}{Received voltage in legitimate Rx node and eavesdropper node vs. different eavesdropper node orientation by changing its angle in a rotational fashion w.r.t. legitimate Rx node position - configuration 4}.}
    \label{fig:config3C2_anglewrtTxnode}
\end{figure}

Fig.~\ref{fig:config3C2_MFD} shows the magnetic flux density based on \textcolor{black}{configuration 4}, where the eavesdropper node had a different orientation with respect to the legitimate Rx node instead of a legitimate Tx node. In this case, we note that the eavesdropper node is farther away, i.e., (6,0,0)ft from the legitimate Tx node when its orientation is $\theta^{Rx-E}=0^\circ$ as shown in Fig.~\ref{subfig:config3C21_MFD}. Therefore, the magnetic flux density $B$ is weaker, and hence, the magnetic flux is minimal. However, the existence of a legitimate Rx node here may change the magnetic field signal strength and act as a waveguide node, which can either tune or de-tune the magnetic field signal at the eavesdropper node. Fig.~\ref{subfig:config3C22_MFD} is a special case because here the orientation of the eavesdropper node and the legitimate Rx node is $\theta^{Rx-E}=90^\circ$, but it can be seen that the angle between the legitimate Tx node and eavesdropper is not equal to 90 degrees, i.e., $\theta^{Tx-E} \neq 0$. Therefore, the eavesdropper node will receive some magnetic field at this orientation, unlike in the case of Fig. \ref{subfig:config3C12_MFD}. In Fig. ~\ref{subfig:config3C23_MFD}, the position of the eavesdropper node is between the legitimate nodes Tx and Rx due to the orientation angle of \textcolor{black}{$\theta^{Rx-E}=150^\circ$. Although the eavesdropper node is close to the legitimate Tx, however, the angle of magnetic field lines is parallel to the eavesdropper node, and consequently, will receive a lower magnetic flux density and magnetic flux.}

Fig.~\ref{fig:config3C2_anglewrtTxnode} shows the voltage received at the legitimate Rx and eavesdropper nodes based on \textcolor{black}{configuration 4}. It can be seen that the eavesdropper node receives the maximum voltage at $\theta^{Rx-E}=180^\circ$, and hence can get the maximum information from the legitimate communication. \textcolor{black}{The minimum received voltage can be observed at $\theta^{Rx-E}=0^\circ$, which keep steadily increasing till $\theta^{Rx-E}=120^\circ$, and then linear increase in voltage is observed from $\theta^{Rx-E}=120^\circ$ to $\theta^{Rx-E}=150^\circ$. In the case of a legitimate Rx node, a minimal to no change can be observed against the angles of the eavesdropper node except in $\theta^{Rx-E}=180^\circ$.}

\subsubsection{\textcolor{black}{Configuration 5}}
\label{sec:OConfig3}
In this configuration, the legitimate nodes Tx and Rx are kept fixed, 2ft apart, at positions (0,0,0)ft and (2,0,0)ft (i.e., $d_{L}^{Tx-E}=2ft$). The eavesdropper node is initially placed 2ft away from the legitimate Rx node at position (4,0,0)ft (i.e., $d^{Rx-E}=2ft$), where the angle between them is $\theta^{Rx-E}=0^\circ$. Then the angle of the eavesdropper node is changed in a rotational fashion at an interval of $\theta=15^\circ$ from $0^\circ$ to $180^\circ$ around its own origin, as shown in Fig.\ref{subfig:config3C33_setup}. 

  \begin{figure*}[h!]
 \centering
    \subfigure[\label{subfig:config3C31_MFD}]{\includegraphics[width=0.30\textwidth]{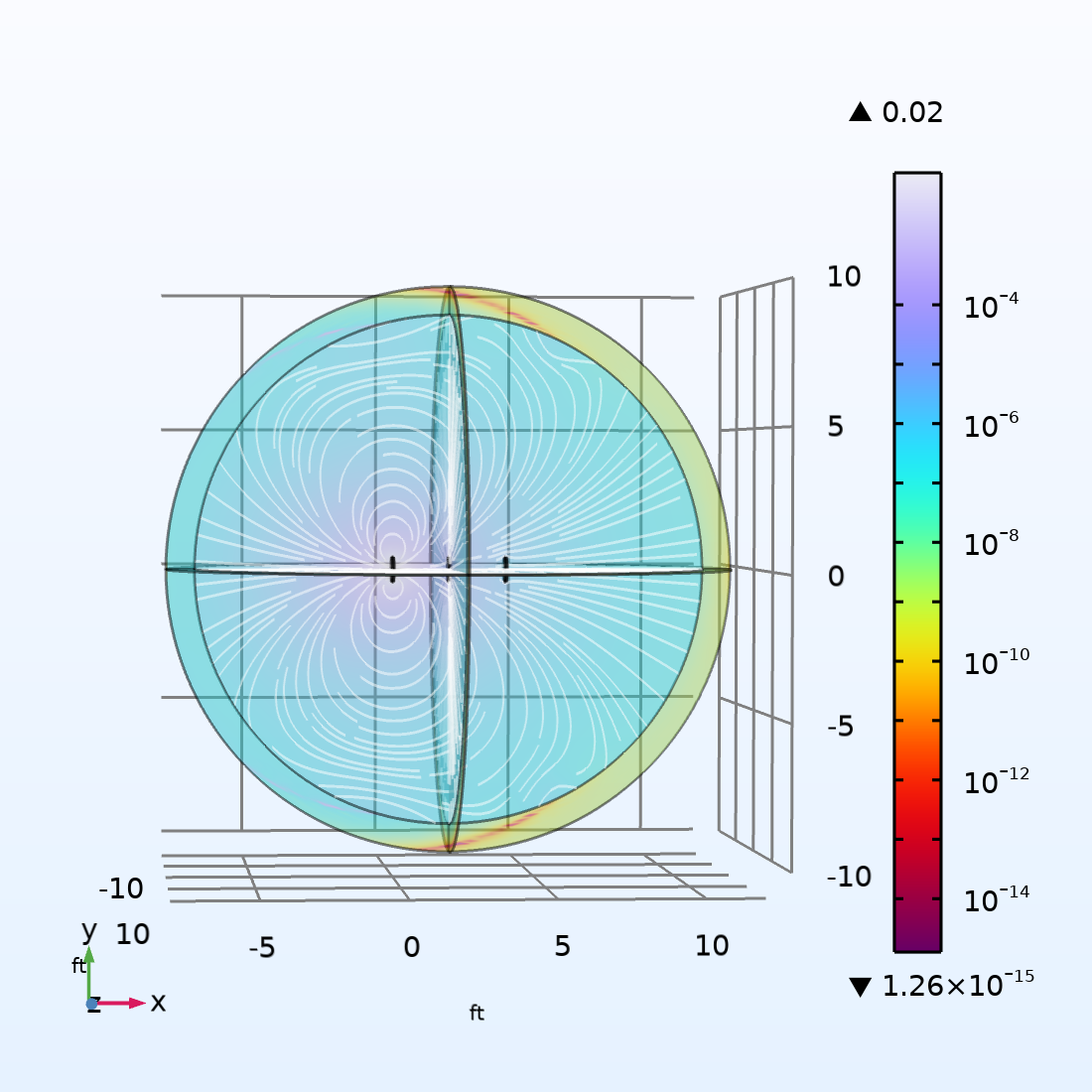}}
    \subfigure[\label{subfig:config3C32_MFD}]{\includegraphics[width=0.30\textwidth]{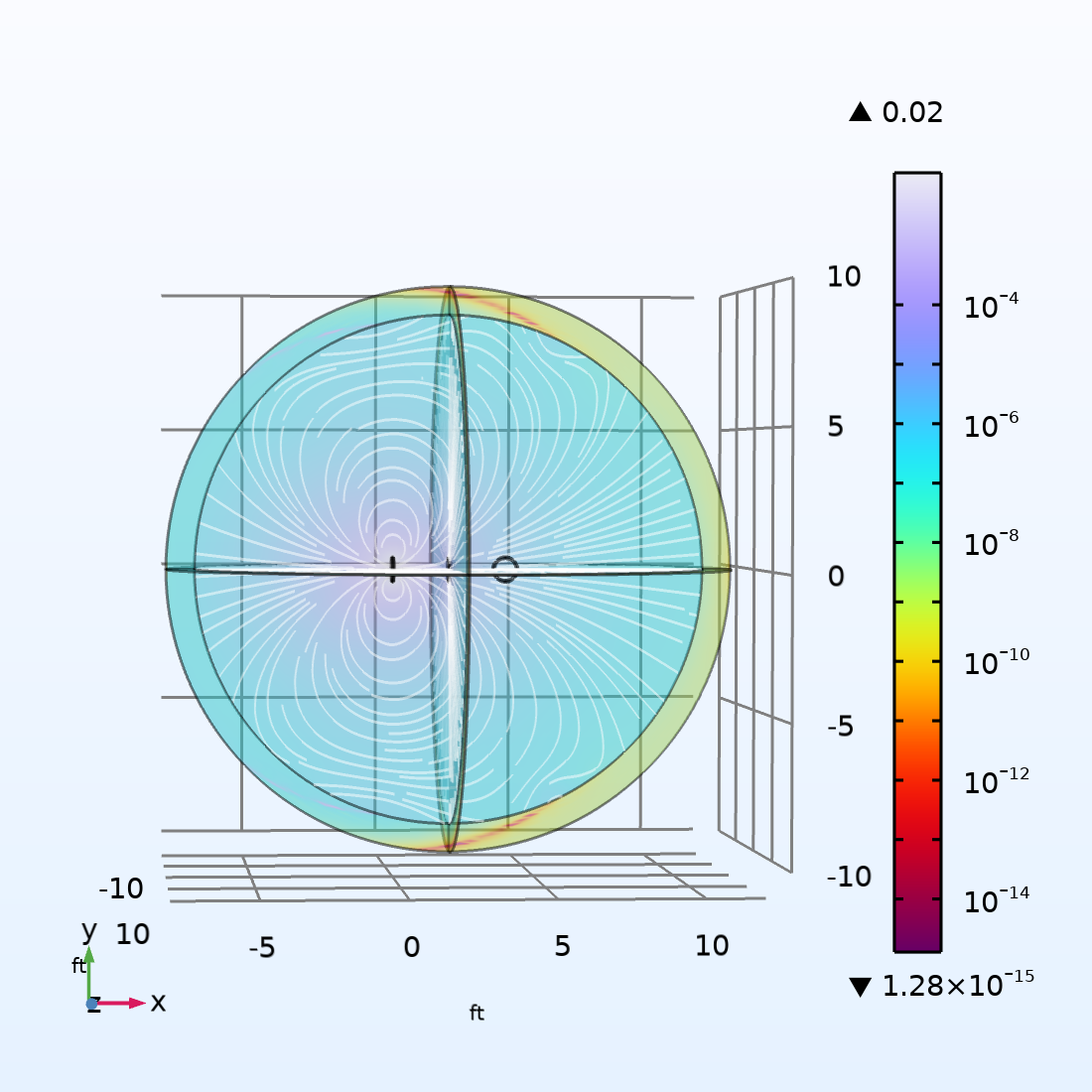}} 
    \subfigure[\label{subfig:config3C33_MFD}]{\includegraphics[width=0.30\textwidth]{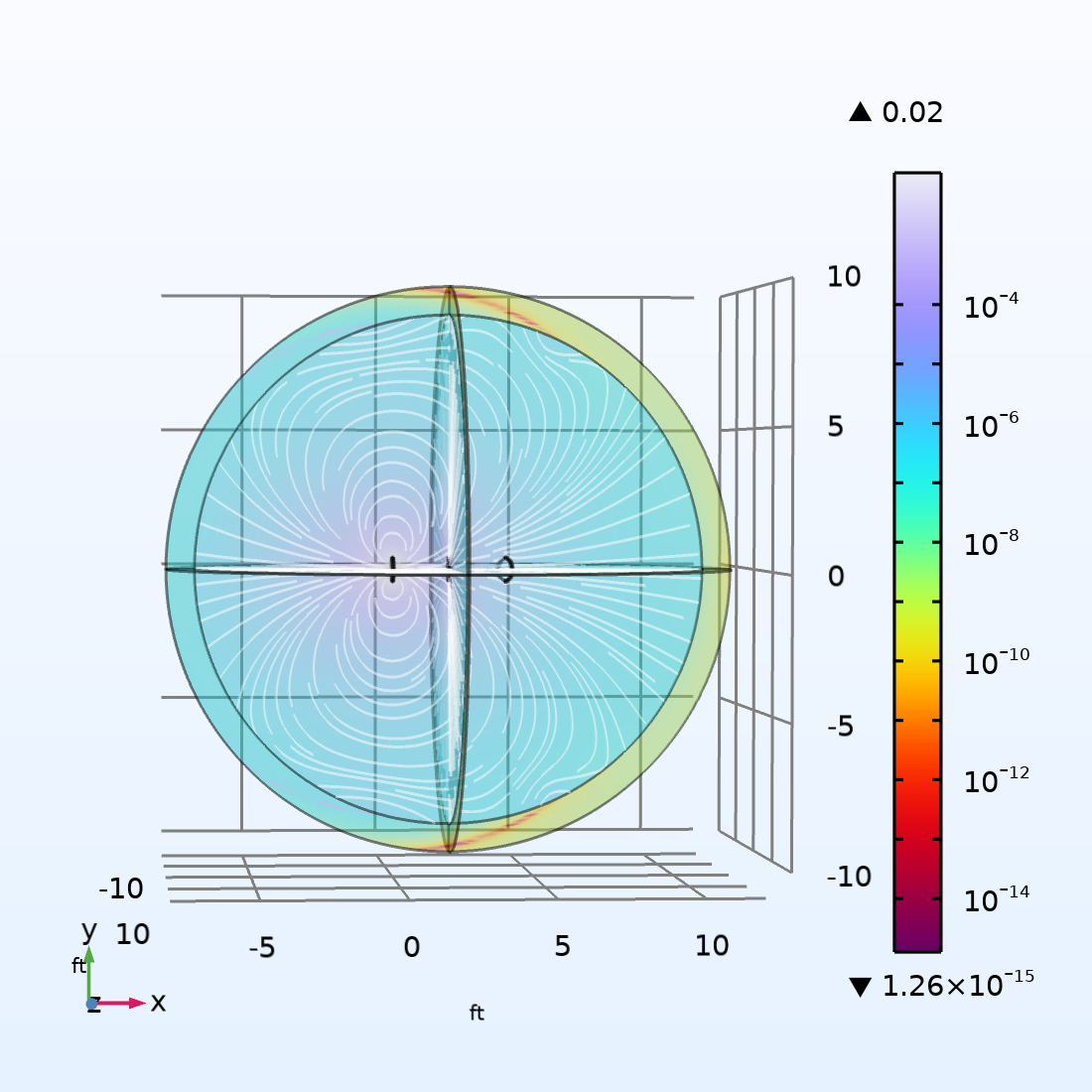}} 
 \caption{Magnetic flux density norm in $T$ with respect to different eavesdropper node angle in the case of configuration 5 when: (\textbf{a}) $\theta^{E}=0^{\circ}$, (\textbf{b}) $\theta^{E}=90^{\circ}$, and (\textbf{c}) \textcolor{black}{ $\theta^{E}=150^{\circ}$.}}
 \label{fig:config3C3_MFD}
 \end{figure*}

   \begin{figure}
    \centering
    \includegraphics[scale=0.60]{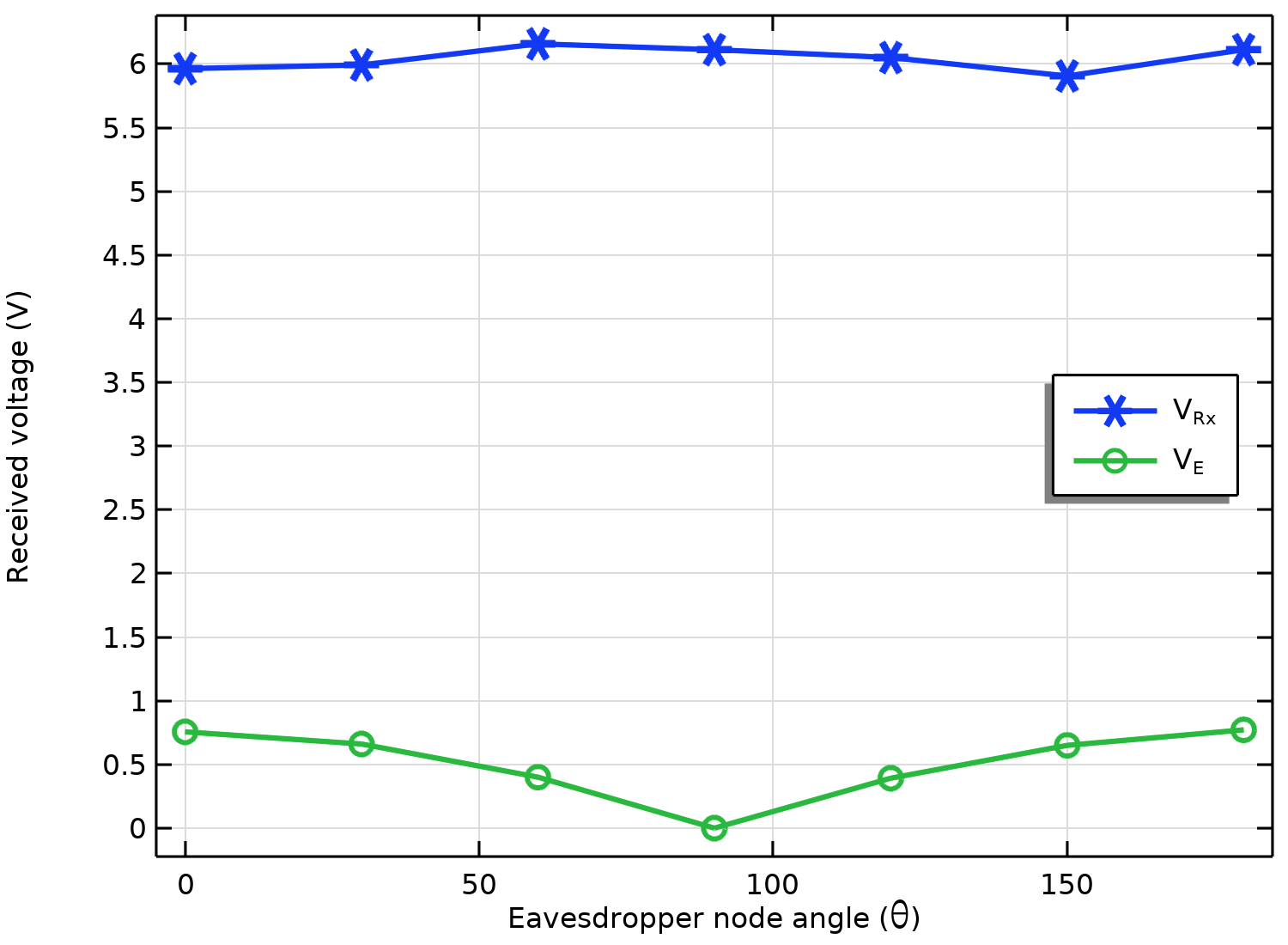}
    \caption{\textcolor{black}{Received voltage in legitimate Rx node and eavesdropper node vs. different eavesdropper node orientation by changing its angle in a rotational fashion w.r.t. its own origin - configuration 5.}}
    \label{fig:config3C3_anglewrtTxnode}
\end{figure}

Fig.~\ref{fig:config3C3_MFD} shows the magnetic flux density in \textcolor{black}{configuration 5}, where the eavesdropper node changes its coil orientation around its own origin. In this configuration, the orientations of $\theta^{Rx-E}=0^\circ$ and $\theta^{Rx-E}=180^\circ$ will have similar results since, in both cases, the eavesdropper node faces the legitimate Tx and Rx nodes. \textcolor{black}{When $\theta^{Rx-E}=150^\circ$, the magnetic flux is lower, since the magnetic field lines and the eavesdropper coil's orientation are nearly parallel to each other.} However, when the orientation of the eavesdropper node is orthogonal, i.e. $\theta^{Rx-E}=90^\circ$ to the legitimate Tx and Rx nodes, it will have zero magnetic flux. Therefore, no magnetic field strength can be received at the eavesdropper node, and consequently, no information can be eavesdropped from the ongoing legitimate MI communication. 

The results of the received voltage at the legitimate Rx and eavesdropper nodes based on \textcolor{black}{configuration 5} are shown in Fig.~\ref{fig:config3C3_anglewrtTxnode}, where the eavesdropper node changes orientation around its own origin. It can be seen from the figure that the overall voltage received by the eavesdropper node at this position is minimal. However, the eavesdropper node can receive maximum voltage at $\theta^{Rx-E}=0^\circ$ and $\theta^{Rx-E}=180^\circ$. However, the received voltage decreases at other orientations and approaches zero received voltage at $\theta^{Rx-E}=90^\circ$. On the other hand, slight changes in the received voltages of the legitimate Rx node can be seen against the orientation of the eavesdropper node.


In conclusion, the position of the eavesdropper node and its coil orientation play a significant role in the eavesdropping of legitimate ongoing MI communication. From the results based on Sections~\ref{sec:sim-Eposition} and \ref{sec:sim-Eorientation}, it can be stated that the eavesdropper node can obtain maximum information from an ongoing legitimate MI communication when the eavesdropper is aligned with and located close to the legitimate position of the Tx node. On the other hand, if the orientation of the eavesdropper node is orthogonal to the Tx node, then it will receive zero information no matter how close it is to the Tx node or the Rx node. In addition, the legitimate node may also sense malicious activities in some eavesdropper position and orientation. However, the legitimate node can sense the change in magnetic flux or received voltage and apply countermeasures to jam the eavesdropper node.

\subsubsection{Secrecy Capacity Evaluation}
\label{sec:sim-ESC}
In this section, we develop a FEM simulation model of an eavesdropping attack and consider secrecy capacity as the performance metric for evaluation. In this setup, the legitimate Tx node is kept fixed at (0,0,0)ft, while the legitimate Rx node is initially placed at (0.5,0,0)ft and then moved away from the Tx node till (4,0,0)ft at an interval of 0.5ft each time. The eavesdropper node is placed initially at (0,4.5,0)ft and moved away till (0,7.5,0)ft at an interval of 1ft each time. 

\begin{figure}
    \centering
    \includegraphics[scale=0.60]{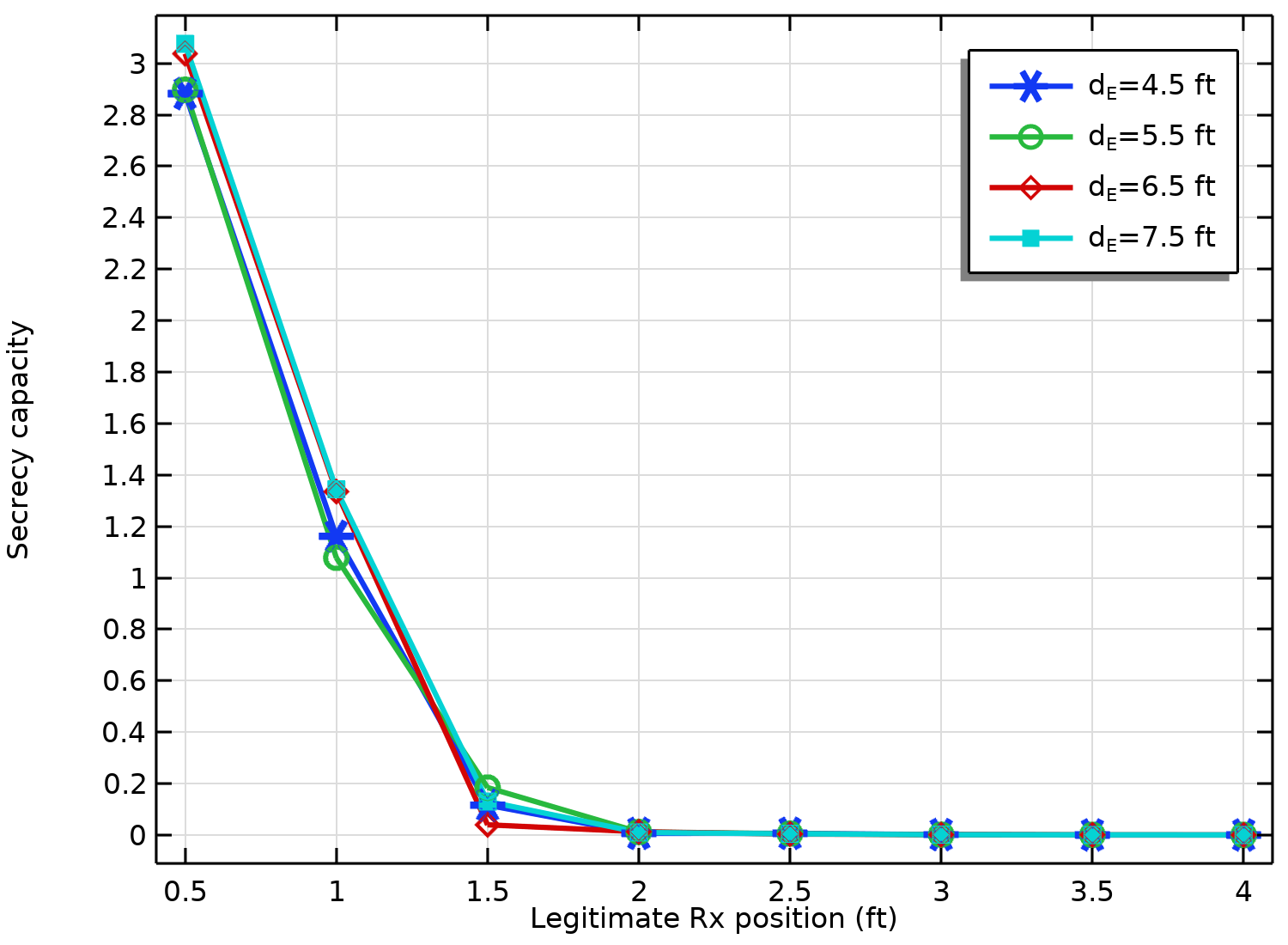}
    \caption{Secrecy capacity vs. legitimate node position under different eavesdropper node positions.}
    \label{fig:sc-sim}
\end{figure}

Fig.~\ref{fig:sc-sim} shows the secrecy capacity versus the legitimate position of the Rx node for different positions of the eavesdropper node. The maximum secrecy capacity can be observed in Fig.~\ref{fig:sc-sim} when the eavesdropper position is (0,7.5,0)ft and the minimum secrecy capacity can be seen when the eavesdropper node approaches the legitimate Tx node, i.e., (0,4.5,0)ft.  Because, in the considered setup for evaluating the secrecy capacity, the legitimate Tx and Rx nodes are kept fixed and only the eavesdropper node is moved away from the legitimate Tx node in parallel. When the eavesdropper node position is far from the legitimate Tx node, then it receives less voltage, and therefore, the second term in \eqref{eq:SC} is less prominent than the first term. Consequently, the secrecy capacity will be maximum and vice versa. To conclude, higher secrecy capacity can be achieved when the eavesdropper node is not closer to the legitimate Tx node than the legitimate Rx node. In addition, better secrecy capacity can be achieved when the legitimate Tx and eavesdropper nodes are not aligned to each other, especially when they are orthogonal to each other. 

\subsection{\textcolor{black}{Experiments-based Study}}
\label{sec:exp}
\begin{figure}[]
    \centering
    \includegraphics[width=\linewidth, height=4cm]{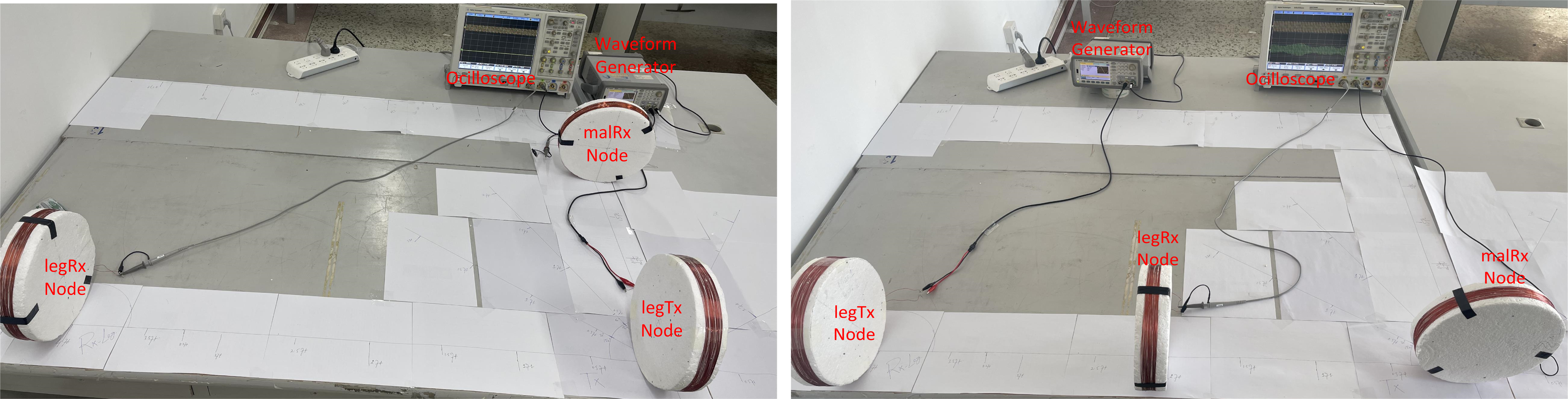}
    \caption{A glance of an experimental lab setup.}
    \label{fig:expsetup}
\end{figure}

We developed a lab experiment to study eavesdropping in MI communication, as shown in Fig. \ref{fig:expsetup}. To compare the results of the FEM simulation and the experimental test setup, we used similar parameters and the same configurations as described in Section \ref{sec:sim-results}. 

In the lab setup, the parameters used in the simulation setup (listed in Table.~\ref{tab:sim-tab}) are kept the same, except for the coils' inductance and capacitance. The capacitance and inductance of the legitimate Tx, Rx, and eavesdropper node are given in Table~\ref{tab:exp-tab}. The legitimate Tx is excited with a sinusoidal voltage signal of $10V$ from the waveform generator. The signals from the legitimate Rx node and the eavesdropper node are recorded from the oscilloscope. 

\begin{table}[]
    \centering
    \caption{Capacitors and coils inductance value used in the experimental setup.}
    \label{tab:exp-tab}
    \begin{tabular}{|c|c|c|c|}
    \hline
         & Tx node & Rx node & Eavesdropper node\\
     \hline  
     Capacitance    & 5.62 $nF$  &  5.62$nF$ & 5.62 $nF$ \\
     \hline    
      Inductance   & 449 $\mu H$ & 447 $\mu H$ & 446 $\mu H$ \\
    \hline   
    \end{tabular}
\end{table}

\begin{figure}[h]
 \centering   \subfigure[\label{subfig:config1Vexp}]{\includegraphics[scale=0.55]{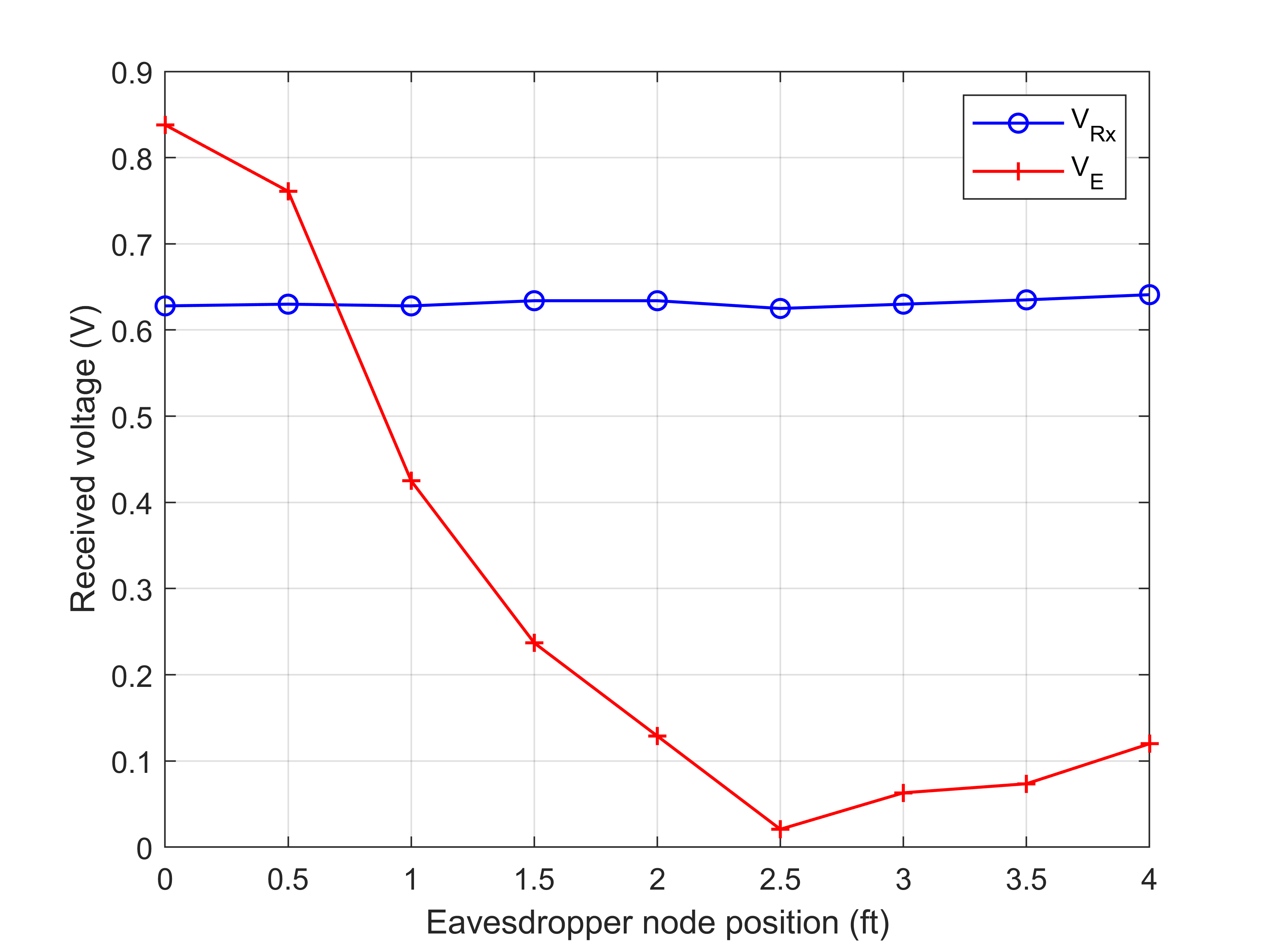}} 
  \subfigure[\label{subfig:config2Vexp}]{\includegraphics[scale=0.45]{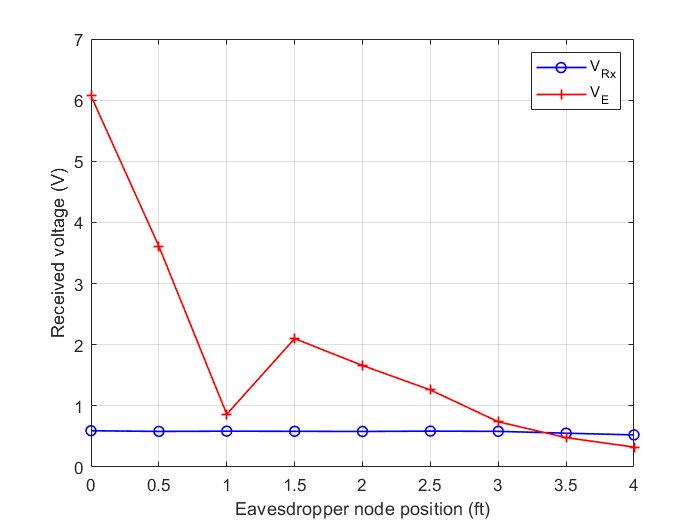}}
   \caption{Experimental results based on eavesdropper node position with respect to legitimate Tx and Rx positions: (\textbf{a}) Received voltage vs. eavesdropper node position under Configuration 1 and (\textbf{b}) Received voltage vs. eavesdropper node position under configuration 2.}
 \label{fig:expEavesPosition}
 \end{figure}

\begin{figure}[]
 \centering
    \subfigure[\label{subfig:config3C1exp_anglewrtTxnode}]{\includegraphics[scale=0.55]{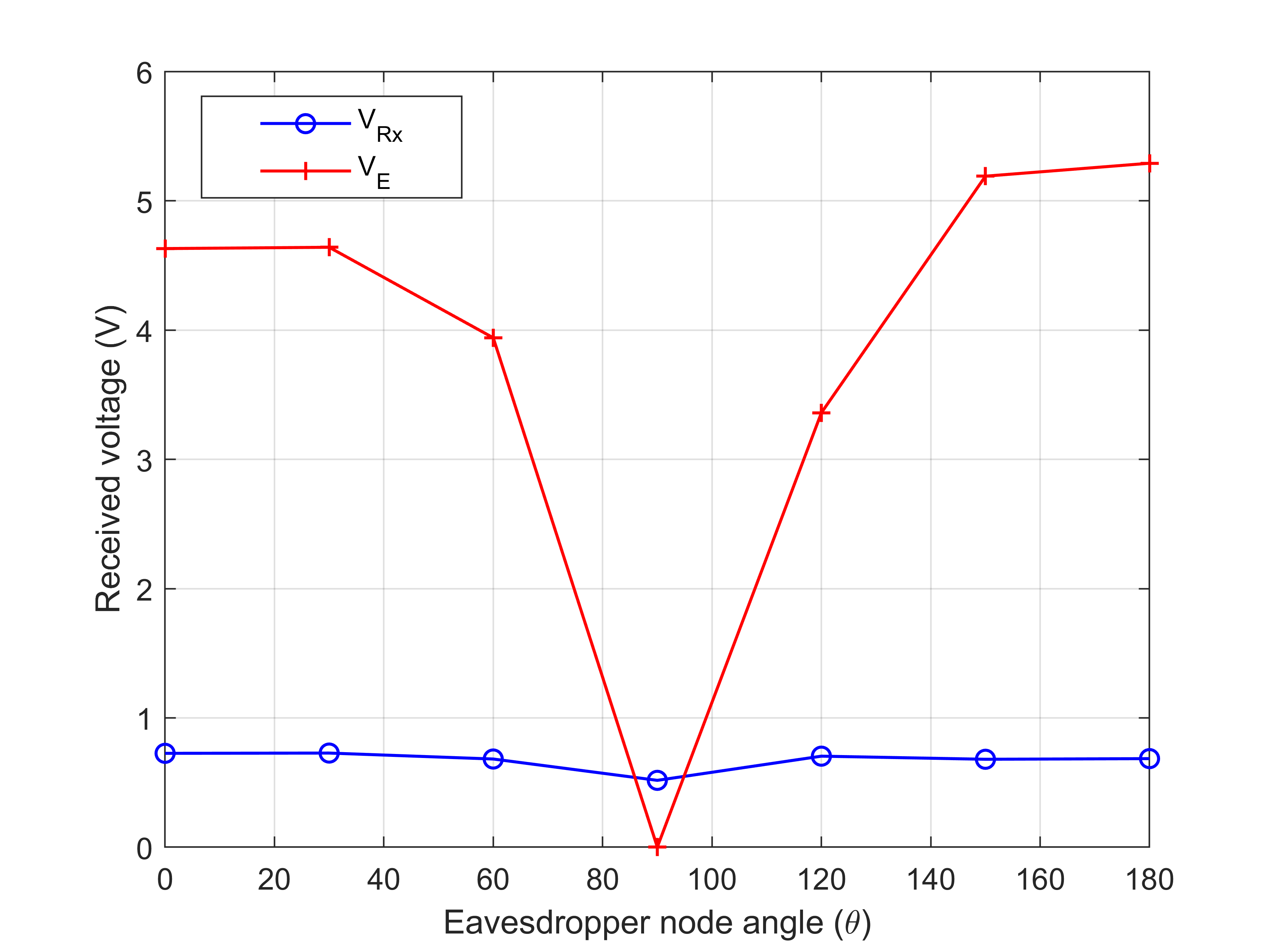}} 
  \subfigure[\label{subfig:config3C2exp_anglewrtTxnode}]{\includegraphics[scale=0.55]{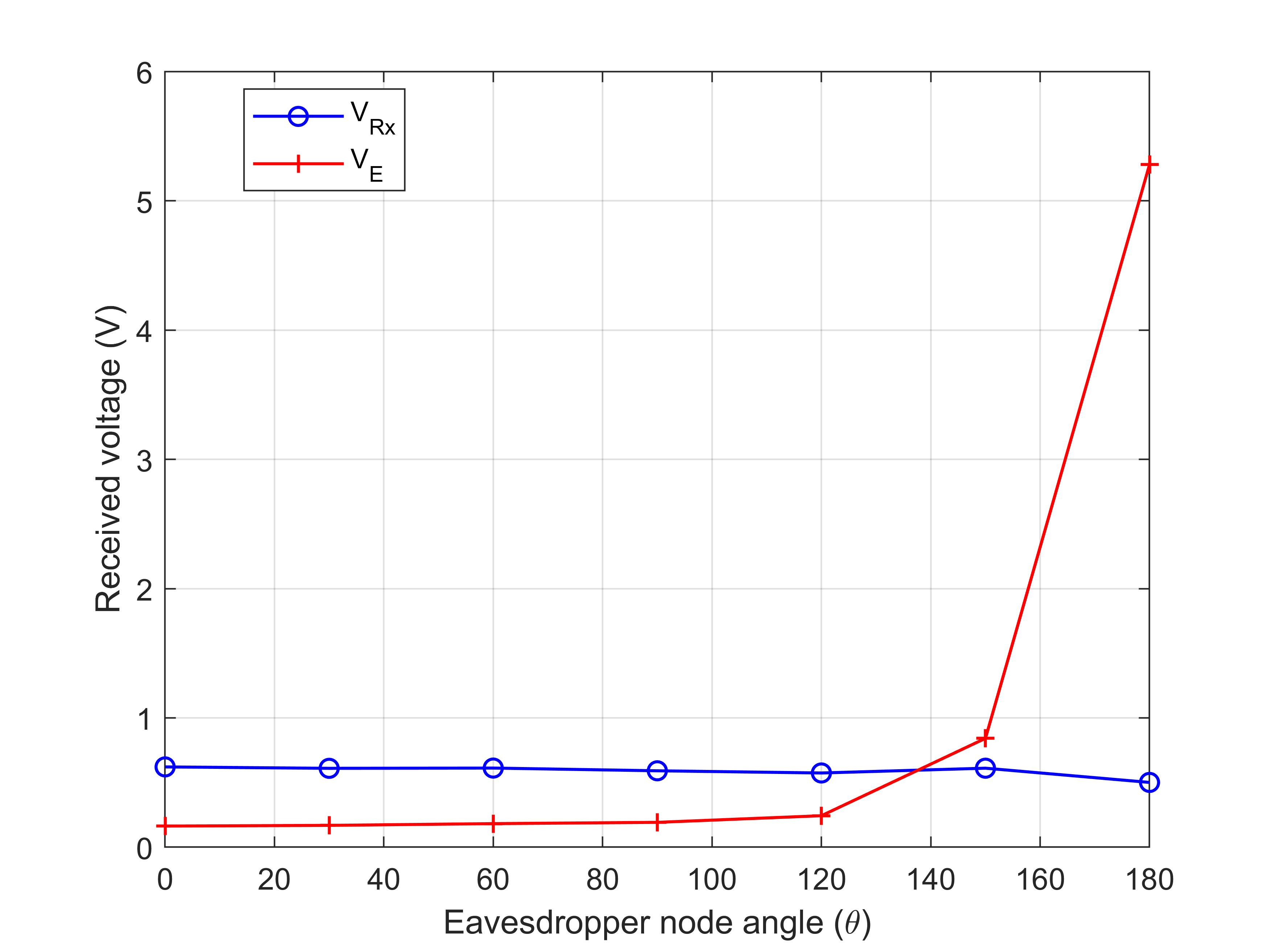}}
   \subfigure[\label{subfig:config3C3exp_anglewrtTxnode}]{\includegraphics[scale=0.55]{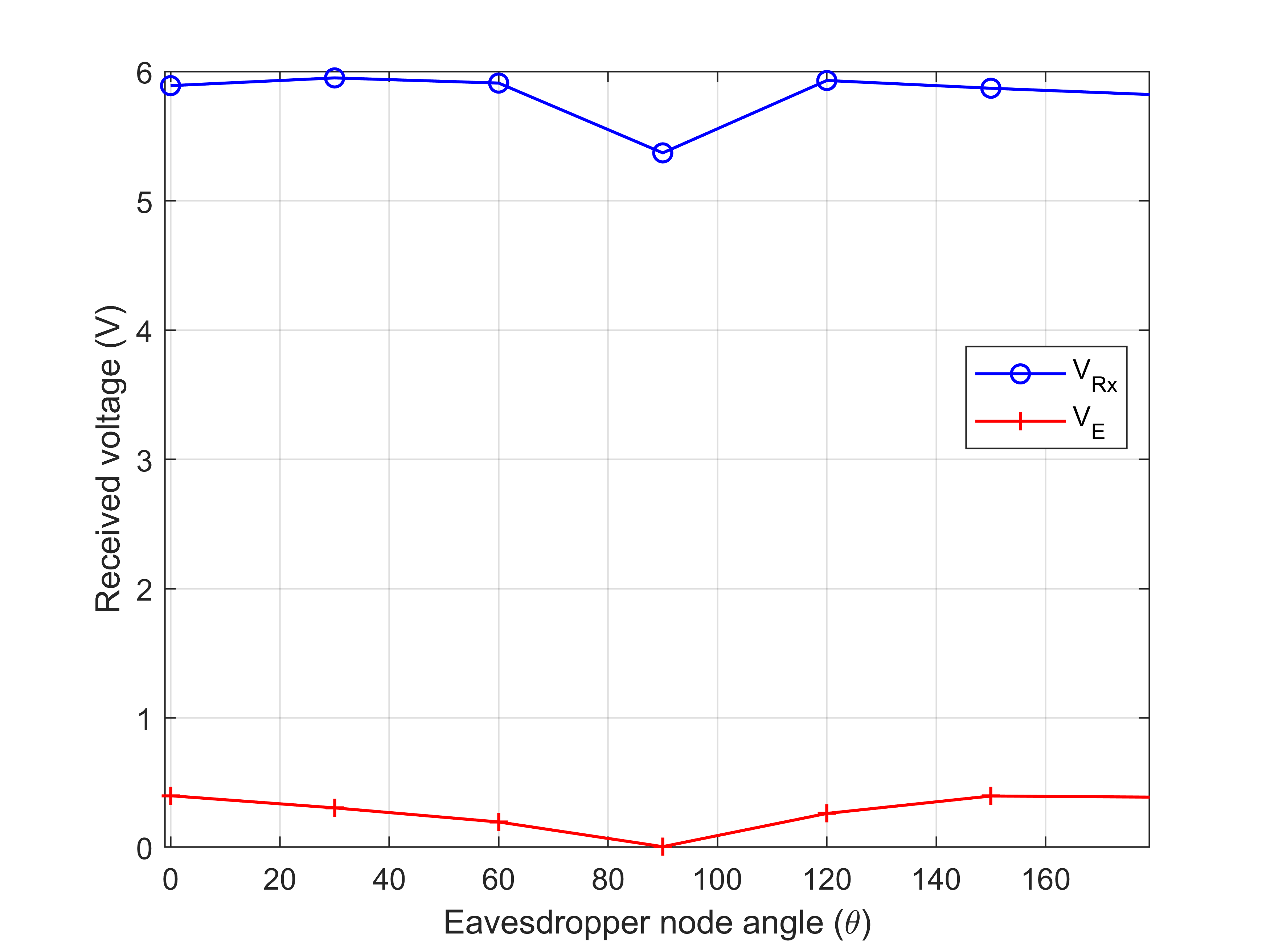}}
 \caption{Experimental setup based received voltage in legitimate Rx node and eavesdropper node vs. different eavesdropper node orientation by changing its angle (\textbf{a}) in a rotational fashion w.r.t. legitimate Tx node position - \textcolor{black}{configuration 3}, (\textbf{b}) in a rotational fashion w.r.t. legitimate Rx node position - \textcolor{black}{configuration 4}, and (\textbf{c}) in a rotational fashion w.r.t. its own origin - \textcolor{black}{configuration 5}.}
 \label{fig:config3Vexp}
 \end{figure}

Fig. \ref{fig:expEavesPosition} shows experimental results of received voltage versus eavesdropper node position (discussed in section \ref{sec:sim-Eposition}) under configurations 1 and 2, respectively (shown in Fig.\ref{fig:config1and2}). Fig.\ref{subfig:config1Vexp} shows the voltage received at the legitimate Rx and eavesdropper nodes vs. different positions of the eavesdropper nodes based on the configuration 1 setup. A similar trend can be seen here compared to the simulation results shown in Fig. \ref{subfig:config1Vsim}. The received voltage at the eavesdropper node changes based on its different position with respect to the legitimate nodes. Similarly, minimal changes in the received voltage of legitimate Rx nodes can be observed against different eavesdropper positions.

The results of the experimental setup based on configuration 2 discussed in Section~\ref{sec:sim-Eposition} are shown in Fig.~\ref{subfig:config2Vexp}. In this configuration, the eavesdropper node approaches the Tx node, and the maximum voltage is received at the eavesdropper node at position (0,1.5,0)ft. While a change in the received voltage of the eavesdropper node can be observed at different positions of the eavesdropper node between legitimate Tx and Rx nodes, minimal to no change can be seen in the legitimate node voltage received. To summarize the results of Fig.~\ref{fig:expEavesPosition}, the eavesdropper node may obtain maximum or minimum information from the ongoing legitimate communication based on its position in the vicinity of the ongoing legitimate MI communication. 

Fig.~\ref{fig:config3Vexp} shows the voltages received at the legitimate Rx and eavesdropper nodes vs. different angles of the eavesdropper nodes based on the experimental lab setup. The subfigure in Fig.~\ref{fig:config3Vexp} shows the experimental results for the three different configurations as discussed in Section~\ref{sec:sim-Eorientation}. The difference between the simulation setup and the experimental setup is that the angle of the eavesdropper node changes at an interval of $\theta=30^\circ$ in the case of the experimental setup. From Fig.~\ref{subfig:config3C1exp_anglewrtTxnode}, it can be seen that the eavesdropper node can receive maximum voltage when the Tx node and the eavesdropper node are facing each other, i.e. $\theta^{Tx-E}=0^\circ$ or $\theta^{Tx-E}=180^\circ$, while the eavesdropper node receives zero voltage at $\theta^{Tx-E}=90^\circ$. On the other hand, the voltage received in the legitimate Rx node shows a significant change at $\theta^{Tx-E}=90^\circ$, otherwise a minimal to no change can be observed. From Fig.~\ref{subfig:config3C2exp_anglewrtTxnode}, it can be seen that the eavesdropper node receives minimal to no voltage until the angle of the eavesdropper node is greater than $120^\circ$. In this case, minimal to no change in voltage can be observed at the legitimate Rx node. In Figure~\ref{subfig:config3C3exp_anglewrtTxnode}, it can be seen that the voltage received in the eavesdropper node is generally minimal, while it approaches zero at an angle of $\theta^{Rx-E}=90^\circ$. The voltage received at the legitimate Rx node shows slight changes, specifically at $\theta^{Rx-E}=90^\circ$. 

In summary, the results achieved from the experimental tests exhibit trends that are almost similar for each configuration to those obtained in the FEM-based simulation results. Results from both the FEM simulation and experimental tests verify that the eavesdropper can eavesdrop on the ongoing legitimate MI communication, but this depends on the eavesdropper's position and orientation with respect to legitimate nodes. Furthermore, legitimate nodes may detect malicious activities in the vicinity due to changes in the MI signal strength/received voltage at the legitimate Rx node. 


\section{Conclusion}
\label{sec:conclusion}
In this paper we present a detailed investigation of the vulnerability of underwater MI communication to eavesdropping attacks. We consider a scenario with three MI nodes: legitimate Tx and Rx nodes, and an eavesdropper node, and answer the following questions: 1) can an eavesdropper eavesdrop on legitimate MI communication and 2) can legitimate nodes sense the malicious activity of the eavesdropper node. For this, we conducted FEM simulations and laboratory experiments. The results show that underwater MI communication systems are indeed vulnerable to eavesdropping attacks, showing that an eavesdropper can eavesdrop on underwater MI-based legitimate communication, mainly based on the position and orientation of the eavesdropper node with respect to legitimate nodes. The results also show that legitimate nodes might be able to detect malicious activities based on the eavesdropper's position and orientation. 


Potential future extensions to this work include: developing a systematic mechanism that can detect eavesdropping from variations in received signals at legitimate coil reception, enhancing the secrecy capacity of the system through resource optimization, and studying the impact of multiple malicious coils present in the surroundings on the information exchange of legitimate coils. 


\bibliographystyle{IEEEtran}
\bibliography{references}

\vfill\break

\end{document}